\documentclass[twocolumn,showpacs,preprintnumbers,amsmath,amssymb,amsfonts,final,a4paper,aps,citeautoscript,footinbib,pra,superscriptaddress]{revtex4-1}
\usepackage[pdftex]{graphicx,color}
\usepackage{dcolumn}
\usepackage{bm}
\usepackage[latin1]{inputenc}
\usepackage{mathrsfs}
\usepackage{epstopdf}
\usepackage{times}
\usepackage{float}
\usepackage[normalem]{ulem}
\usepackage[sort&compress]{natbib}
\usepackage[colorlinks,bookmarks=true,citecolor=blue,linkcolor=blue,urlcolor=blue, breaklinks=true]{hyperref}
\graphicspath{{./figures/}}

\newcommand{\bra}[1]{\langle #1 |}  
\newcommand{\ket}[1]{| #1 \rangle}  
\newcommand{\braket}[2] {\langle #1 | #2 \rangle}

\newcommand{\be}{\text{e}}

\begin{document}
\title{From coupled-wire construction of quantum Hall states to wave functions and hydrodynamics}


\author{Yukihisa Imamura}
\affiliation{Division of Physics and Astronomy, Graduate School of Science, Kyoto University, 
Kyoto 606-8502, Japan.}
\affiliation{%
 Yukawa Institute for Theoretical Physics, Kyoto University, Kyoto 606-8502, Japan}%
\author{Keisuke Totsuka}%
\affiliation{%
 Yukawa Institute for Theoretical Physics, Kyoto University, Kyoto 606-8502, Japan}%
 \author{T.H. Hansson }%
\affiliation{%
 Yukawa Institute for Theoretical Physics, Kyoto University, Kyoto 606-8502, Japan}%
 \affiliation{%
Department of Physics, Stockholm University, AlbaNova University Center, SE-106 91 Stockholm, Sweden}%
 

\begin{abstract}
In this paper we use a close connection between the coupled wire construction (CWC) of Abelian quantum Hall states 
and the theory of composite bosons to extract the Laughlin wave function and the hydrodynamic effective theory in the bulk, 
including the Wen-Zee topological action, directly from the CWC.  
We show how rotational invariance can be recovered by fine-tuning the interactions.  
A simple recipe is also given to construct general Abelian quantum Hall states described by the multi-component Wen-Zee action.  
\end{abstract}

\maketitle

\section{\label{sec:level1}Introduction}
Topological order \cite{WenText} is one of the most fundamental concepts in modern condensed matter physics.
The history starts with the  discovery of the $\nu = \frac{1}{3}$ fractional quantum Hall effect (FQHE) in the 1980's \cite{TsuiStormerGossard},  
which was essentially understood after Laughlin proposed his famous wave function \cite{Laughlin-83}. Since then, theorists have proposed
a variety of topologically ordered states in two and three dimensions.  
Examples in $2+1$ dimensions are Abelian hierarchical quantum Hall (QH) states \cite{Haldane-FQH-83,Halperin-hierarchy-84,Jain-book-07}, 
non-Abelian QH states \cite{Moore1991,Read1999}, 
and various kinds of spin liquids \cite{Kalmeyer-L-87,Wen-W-Z-89,Wen-spinliq-91,Moessner-S-RVB-01,Kitaev-03,Levin-W-05}.   
Some, but far from all, of these states have strong experimental support.   

Topologically ordered states are featureless and symmetric in the bulk,  and as such they defy the characterization 
of phases by local order parameters in the manner of Landau. Nevertheless, there are field theoretic descriptions.  
It is belived that the topological properties can be encoded in various types of topological field theories, 
the most well known examples being the Chern-Simons theories of the QHE \cite{Blok1990,Wen-Z-92}.  

Going beyond the topological scaling limit \cite{Frohlich-K-91}, but still in the infrared region,  
there are hydrodynamical descriptions that supplement the topological action with higher-order derivative terms.  
These theories typically encode information about collective excitations.  
Yet another type of field theories are those based on statistical transmutation, or ``flux attachment".   
These theories of ``composite'' fermions \cite{Lopez-F-91} or bosons \cite{Zhang-H-K-89},  
which are closely related to various model wave functions, are in principle microscopic, 
but can only be solved using various kinds of mean-field approximations. 

In addition to the various field theories, there are several other ways to describe topologically ordered states in general,  
and quantum Hall states in particular.  Examples of the latter are the approach based on the thin-torus limit \cite{bergholtz2008},  
and the coupled wire construction (CWC) of Kane and coworkers \cite{KaneMukhopadhyayLubensky,Teo-K-14}.   
The aim of this paper is to make a rigorous connection between this last approach and the Chern-Simons field-theory for composite bosons. 

The CWC is quite general, and  has been employed to construct various two- and three-dimensional topological states.   
The list includes  the original work on Abelian and non-Abelian fractional quantum Hall states 
\cite{KaneMukhopadhyayLubensky,Teo-K-14,Fuji-Y-17},  
chiral spin liquid states \cite{Meng2015}, topological insulators and superconductors 
in two- and three dimensions \cite{Sagi2015,Santos-H-G-G-15,Sagi-O-14,Seroussi2014}, 
and the construction of higher-dimensional Abelian topological phases \cite{Iadecola-N-C-M-16}.  
A re-derivation of the periodic table of integer and fractional fermionic topological phases was given in \cite{Neupert2014}.  
In this paper, we shall only consider Abelian QH states, and in particular the Laughlin states.

The essence of the CWC is to build interacting  fermion/boson systems by starting from a collection of parallel ``wires"  in a strong magnetic field,
each one supporting a Luttinger liquid. The wires are then coupled by tunneling interactions, and  there are also forward scattering interactions on each wire that couple right and left moving electrons. Keeping only the intra-wire current-current interactions, 
the system is in the so-called  sliding Luttinger liquid phase \cite{Kivelson-F-E-98,O'Hern-L-T-99,Emery-F-K-L-00,Mukhopadhyay-K-L-SLL-01,Vishwanath-C-01} and remains invariant under {\em independent} global U(1) transformations and translations on  each wire.

A key observation is that  inter-wire interactions can be used to freeze most of the above symmetry.   
Kane {\em et al.} \cite{KaneMukhopadhyayLubensky}, showed that, in the limit of strong inter-wire coupling, and at certain rational filling fractions,
one retains various two-dimensional ground states  whose properties depend 
on the the details of the construction.  
In the simplest case of the Laughlin state, it is sufficient to couple neighboring wires, but in general several wires have to be coupled.   

The CWC gives an intuitive description of the chiral edge states in a way resembling the occurrence of fractional spins at the ends 
of the Affleck-Kennedy-Lieb-Tasaki chain \cite{Affleck-K-L-T-88}, 
or of Majorana modes at the ends of a Kitaev chain \cite{Kitaev-Majorana-01}.  The basic mechanism is that the right 
and left-moving electrons on  neighboring chains pair, leaving a right-moving channel on one side of the sample and a left-moving one on the other.  
One can also show that the number of degenerate ground states are as expected, and that the kink excitations on the wires  indeed  are the
fractionally charged anyons characteristic of FQH liquids.

It is less clear how the CWC, which, by definition, explicitly breaks rotational invariance, will describe other characteristics 
of the Laughlin states, such as its wave function and collective excitations.  
In this paper, we reformulate the CWC in terms of gauge fields, in a way that makes it clear how to reproduce the known bulk properties.  
Not surprisingly, a fine-tuning of a parameter is needed to get an isotropic two-dimensional liquid, but we also show that  
the topological properties do not depend on this.  Concretely, we shall use the bosonic fields in the CWC to define a gauge field  
that can be identified with the statistical gauge field in the Chern-Simons-Ginzburg-Landau (CSGL) theory 
of the Laughlin states \cite{Zhang-H-K-89,Zhang-CSGL-92}.   
Using this, we derive the Laughlin wave function, both in the isotropic and anisotropic cases, and the hydrodynamic effective theory which contains 
the Wen-Zee topological action \cite{Wen-Z-92} as its leading term in a derivative expansion. 

A comment is in order about an interesting recent paper by Fuji and Furusaki \cite{Fuji-F-19}.  
In fact, there is an overlap in the underlying idea between it and this work.  
Rather than directly identifying the composite-boson field in the CWC,  
they explicitly carry out the flux attachment on the wires, and then add an interaction term that stabilizes the superfluid phase of the composite bosons.  
Finally, using a coupled-wire version of the boson-vortex duality transformation \cite{Fisher-L-89}, they arrive at an action that couples 
the dual (vortex) fields, describing the quasiparticles, to a hydrodynamic gauge field with a Chern-Simons action.  
Despite similar looking, this theory differs in some important respects from the hydrodynamic effective theory we shall derive 
below.  
We comment on this in the end of Sec.~\ref{sec:hydrodynamic-WZ}. 
 Also, fluctuations and the wave function, which are main subjects of this paper, are not discussed in Ref.~\cite{Fuji-F-19}.  

The paper is organized as follows.  
We begin by briefly reviewing the CWC in Sec.~\ref{sec:CWC-for-Laughlin} and  show how to  identify the physically 
inequivalent ground states.  
In Sec.~\ref{sec:gauge-from-CWC}, we define a gauge field which can be interpreted as a statistical gauge field 
and explain the  connection to the CSGL theory.  
Section \ref{sec:Laughlin-wf} contains the derivation of the Laughlin wave function starting from our effective Hamiltonian, as well as 
a discussion of the effects of the spatial anisotropy which is naturally present in the coupled wires.   
In Sec.~\ref{sec:hydrodynamic-WZ}, we derive the hydrodynamical theory, that contains the topological Wen-Zee action, 
from the effective action obtained in the previous section and show that the Kohn mode is correctly reproduced. 
The generalization to arbitrary Abelian states is given in Sec.~\ref{sec:general-TQO-2D}, and we end with a concluding section. 
 Some technicalities are summarized in two Appendices.  

\section{Laughlin state in Coupled Wire Construction}
\label{sec:CWC-for-Laughlin}
The Laughlin state is the paradigmatic example of topological states of matter (see, e.g., Refs.~\onlinecite{Jain-book-07,Hansson-H-S-V-17} 
for reviews of the Laughlin and other quantum Hall states).  
The defining properties of the $\nu=1/m$ state are encoded in the Wen-Zee topological action which is 
a  level-$m$ U(1) Chern-Simons gauge theory \cite{Blok1990,Lopez-F-91}.  
The corresponding edge theory is a chiral Luttinger liquid, which is a chiral boson compactified on a circle 
with radius $\sqrt m$ \cite{Wen-CLL-90}.  
We now review the CWC for the Laughlin state following the original work by Kane {\em et al.} \cite{KaneMukhopadhyayLubensky}. 

\subsection{From uncoupled to  to coupled wires} \label{IIA}
\begin{figure}
\includegraphics[width=80mm]{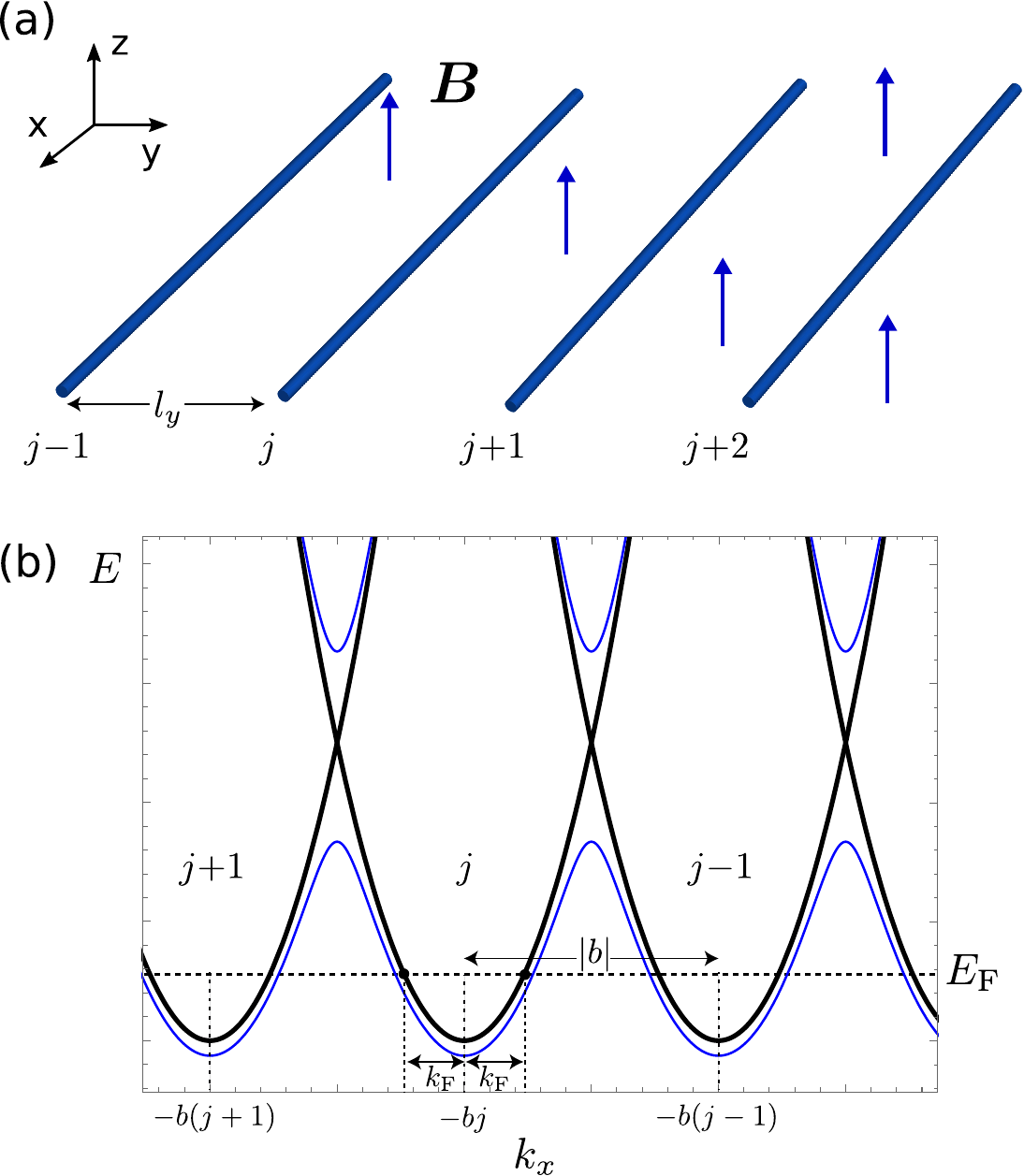} 
\caption{(Color online) %
(a)
An array of coupled wires in a perpendicular magnetic field.
(b)
The energy band structure of the coupled wires.
In the limit of decoupled wires, a perpendicular magnetic field in the Landau gauge shifts 
the quadratic dispersion relations of the individual wires (labeled by $j$) in a $j$-dependent manner: $k \to k + bj$ (thick curves).  
When $2k_{\text{F}}=|b|$ ($\nu=1$), 
the single-particle hopping between the adjacent wires opens the gaps at Fermi points 
and the anisotropic Landau levels are formed (thin blue curves). 
\label{fig:wires}}
\end{figure}

Consider a set of parallell one-dimensional wires running along the $x$ direction and stacked in the $y$ direction with a distance $l_y$ 
[see Fig.~\ref{fig:wires}(a)]. 
We then apply a strong magnetic field $B\, (>0)$ which is perpendicular to the $xy$ plane and is given by 
the gauge potential $\mathbf{A}=(-By,0)$ (Landau gauge).   
We consider non-relativistic spinless fermions $c_{j}(x)$ (with mass $M_{\text{e}}$) moving on the wire $j$  
with the dispersion relation $E_{j}(k_x) = (k_x + bj)^2 / 2M_{\text{e}}$, where $b=e B l_y /\hbar$ [see Fig.~\ref{fig:wires}(b); 
$e<0$ for electrons].  
When these bands are filled up to the Fermi momentum $k_\text{F}$ [see Fig.~\ref{fig:wires}(b)], 
the fermion density on each wire is given by $\bar{\rho}^{\text{1D}} = k_{\text{F}} /\pi$, 
and this gives the filling fraction: $\nu=2k_\text{F} / |b|$.  
When $\nu=1$, the Fermi points are located exactly at the lowest crossing points and 
the inclusion of the single-particle hopping $c^{\dagger}_{j+1}(x)c_{j}(x)+(\text{H.c.})$ among the neighboring wires opens a band gap there, 
which is identified with one of the gaps separating the (quasi-one-dimensional) Landau levels as is seen 
in the thin curves in Fig.~\ref{fig:wires}(b).   

At a filling fraction, $\nu <1$,  which is relative to the filled Landau levels, the single-particle hopping never opens a gap at the Fermi points 
and we need interactions to make the system gapped.  
To develop a systematic approach, we linearize  the low-energy dispersion around the  Fermi points 
$k_{x}^{\text{R}}(j) = - bj+k_{\text{F}}$ and  $k_{x}^{\text{L}}(j) = - bj- k_{\text{F}}$ to obtain two Dirac fermions 
$R_j$, and $L_j$ respectively, which are related to the original spin less fermion $c_{j}(x)$ as: 
\begin{equation}
c_{j}(x) \approx \be^{i k_{x}^{\text{R}}(j)x} R_{j}(x) + \be^{i k_{x}^{\text{L}}(j)x} L_{j}(x)  \; .  
\label{eqn:spinless-F-by-R-L}
\end{equation}
The low-energy effective Hamiltonian for the $j$-th wire is given by that of massless Dirac fermion:
\begin{equation}
{\cal H}^{(j)}_{\text{Dirac}} = v^{0}_{\text{F}}  \int\! dx \left[- i R^{\dagger}_{j} \partial_{x} R_{j}  
+ i L^{\dagger}_{j} \partial_{x} L_{j} \right]  \; ,
\label{eqn:Dirac-action}
\end{equation}
where the velocity $v^{0}_{\text{F}} =\hbar k_{\text{F}}/M_{\text{e}}$ is common to all wires.  
(To ease the notations, we set $\hbar=1$ in what follows.) 

By Abelian bosonization \cite{Giamarchi-book-04},  
the chiral spinless fermions $R_{j}$ and $L_{j}$ on each wire ($j$) are expressed, at low energies, in terms of the chiral bosons 
$\phi_{j}^{\text{L/R}}$ as  
\begin{equation}
\begin{split}
& R_{j} = \frac{\kappa_{j}}{\sqrt{2\pi a_0}} 
: \exp\left( 2 i  \phi_{j}^{\text{R}}\right) : \, , \; 
L_{j} = \frac{\kappa_{j}}{\sqrt{2\pi a_0}} 
: \exp\left( 2 i \phi_{j}^{\text{L}}\right) :  \; ,
\end{split}
\label{eqn:spinless-boson2fermion}
\end{equation}
where $\{ \kappa_{j} \}$ are the Klein factors necessary to guarantee the anti-commutation among the spinless 
fermions on different wires.   The symbol $: \cdots :$ denotes the normal-ordering necessary to regularize 
the operator-products; to simplify the notations, we suppress it hereafter. 
In terms of the bosonic fields $\Phi_{j}$ (compactified on a circle with radius $1$) 
and their duals $\Theta_{j}$
\begin{equation}
\Phi_{j}= \phi_{j}^{\text{L}}+\phi_{j}^{\text{R}} \; , \quad 
\Theta_{j}=  \phi_{j}^{\text{L}}-\phi_{j}^{\text{R}} \; ,
\label{eqn:PhiTheta2chiral}
\end{equation}
the low-energy effective Hamiltonian for the spinless fermion is
\begin{equation}
{\cal H}^{(j)}_{\text{Dirac}} = \frac{v_{\text{F}}}{2\pi}\int\!dx \sum_{j=1}^{N_{y}}  
\left\{
\frac{1}{K_j}: \!(\partial_{x}\Phi_{j})^{2}\! :
+K_j : \! (\partial_{x}\Theta_{j})^{2}\! :
\right\}  \, ,
\label{eqn:Ham-Dirac}
\end{equation}
where $N_y$ is the number of wires in the stacking ($y$) direction. The bosonic fields satisfy 
\begin{equation}
\begin{split}
& [\Phi_{j}(x),\Phi_{j^{\prime}}(x^{\prime})]=[\Theta_{j}(x),\Theta_{j^{\prime}}(x^{\prime})]=0  \\
& [ \partial_{x}\Phi_{j}(x), \Theta_{j^{\prime}}(x^{\prime})] = -i \pi \delta_{j j^{\prime}} \delta(x-x^{\prime}) \; ,
\end{split} 
\label{eqn:boson-commutation}
\end{equation}
and the Luttinger liquid parameter $K_j$ equals to 1 for free fermions but, in general, $K_j$ and $v_{\text{F}}$ are 
renormalized in the presence of interactions. 
The field $\Theta_{j}$ is related to the particle density [measured from its average 
$\bar{\rho}^{\text{1D}} = k_{\text{F}} /\pi = \nu |b|/(2\pi)$] on the $j$-th wire:
\begin{equation}
\rho^{\text{1D}}_j (x) = \bar{\rho}^{\text{1D}} + \delta \rho^{\text{1D}}_j (x) = 
\bar{\rho}^{\text{1D}} + \partial_{x}\Theta_j(x)/\pi  \; .
\label{eqn:1D-density-by-Theta}
\end{equation}
This relation will be used frequently in the following sections.  

To introduce interactions, we consider an adjacent pair, e.g., $j$ and $(j+1)$.
In order to gap out most of the degrees of freedom and leave gapless 
chiral Luttinger liquids only at the edges, the authors of 
Refs.~\cite{KaneMukhopadhyayLubensky,Teo-K-14} introduced the following, carefully designed, inter-wire interaction 
allowed by U(1) and translation symmetries:
\begin{equation} 
\mathcal{H}_{\text{int}}
= - g \sum_{j}
: \left(L_{j}^{\dagger}R_{j} \right)^{\frac{m-1}{2}} L_{j}^{\dagger} 
\left(L_{j+1}^{\dagger}R_{j+1} \right)^{\frac{m-1}{2}} R_{j+1}   :   \; , 
\label{eqn:inter-wire-int-by-fermion}
\end{equation}
where $m$ is an odd integer for fermions \footnote{%
At this stage, $m$ must be integer as the number of back-scattered fermions must be integer: $(m-1)/2 \in \mathbb{Z}$.  
However, once the interaction is given in terms of bosons $\Phi_j$ and $\Theta_j$ as in Eq.~\eqref{eqn:inter-wire-int-Laughlin}, 
we can think of bosonic problem with even-$m$.}.     

This interaction is made up of two simultaneous $\frac{m-1}{2}$-pled (on-wire) backscattering processes 
on the adjacent wires and inter-wire single-particle hopping (i.e., $\mathcal{H}_{\text{int}}$ consists of 
$m$-particle processes).  According to Eq.~\eqref{eqn:spinless-F-by-R-L}, 
the chiral fermions $R_{j}(x)$ and $L_{j}(x)$ are accompanied, respectively, by the oscillating factors 
$\be^{i k_{x}^{\text{R}}(j)x}$ and $\be^{i k_{x}^{\text{L}}(j)x}$.  Therefore, if we construct the interaction $\mathcal{H}_{\text{int}}$ 
from the original $c_j$-fermions, it must contain the factor $\be^{i(-b+2m k_{\text{F}})x}$.  
Requiring  $\mathcal{H}_{\text{int}}$ to be non-oscillatory (i.e., translationally invariant) fixes the filling fraction $\nu=2k_\text{F} / |b|=1/m$.  

\begin{figure}[htb]
\includegraphics[width=85mm]{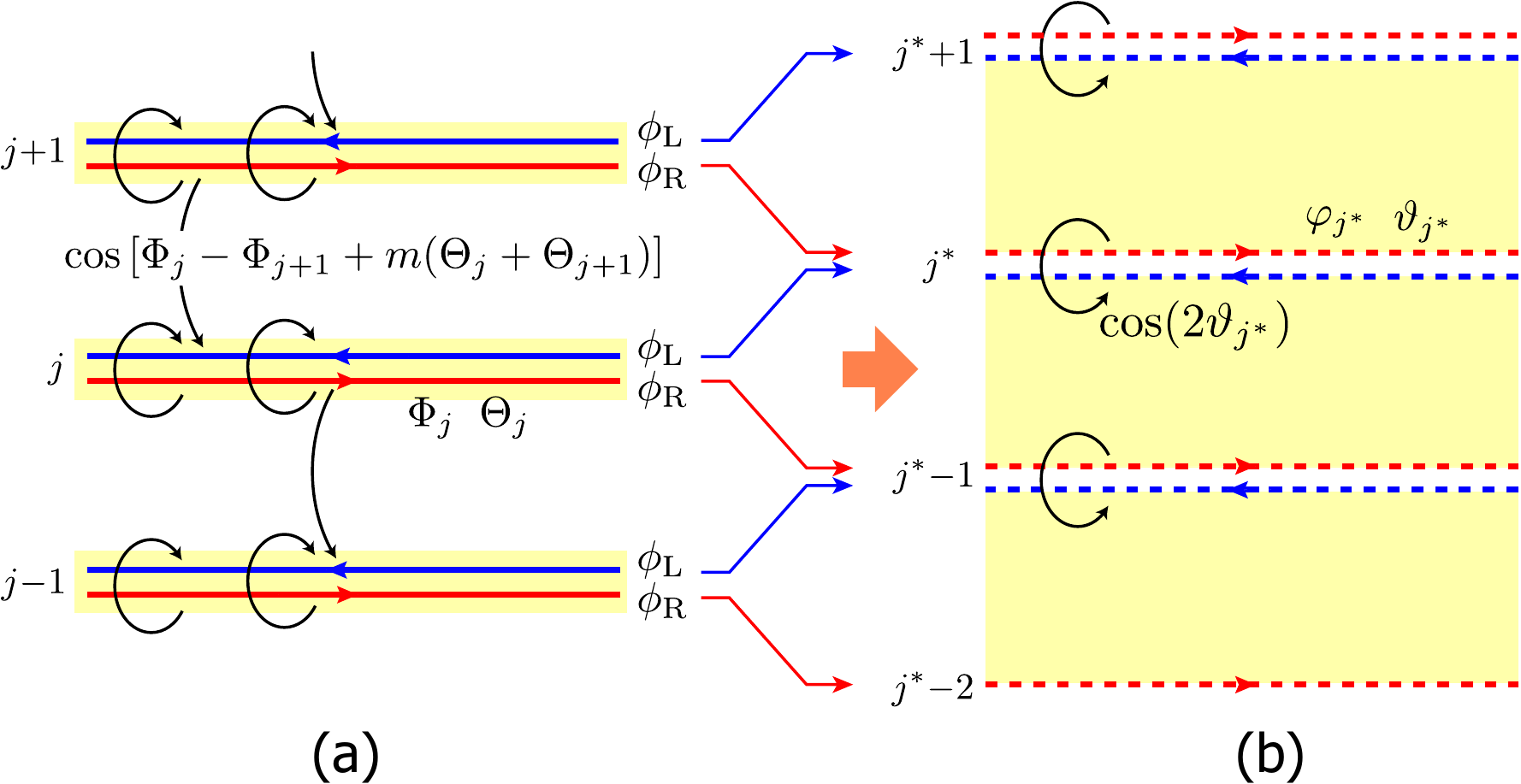} 
\caption{(Color online) %
(a) original wires (located at $j$) on which the original bosons $\{ \Phi_j,\Theta_j \}$ are defined and (b) virtual wires 
(with positions specified by $j^{\ast}$) corresponding to the on-strip fields $\{ \varphi_{j^{\ast}},\vartheta_{j^{\ast}} \}$.  
Dashed lines denote the new set of chiral bosons $\tilde{\phi}^{\text{L/R}}_{j^{\ast}}$,   
with which inter-wire interaction \eqref{eqn:inter-wire-int-Laughlin} 
looks like an ordinary backscattering $\cos(2\vartheta_{j^{\ast}})$.  
\label{fig:wire-vs-strip}}
\end{figure}

In terms of the bosonic fields, $\mathcal{H}_{\text{int}}$ reads as \footnote{
We were not  precise about the ordering of the fermion operators in \eqref{eqn:inter-wire-int-by-fermion}; 
we assume that they are correctly ordered in such a way that, when bosonized, they reproduce \eqref{eqn:inter-wire-int-Laughlin}. 
} 
\begin{multline}
\mathcal{H}_{\text{int}} 
= - g \sum_{j} \int dx : \cos \left[ \Phi_{j} - \Phi_{j+1} + m (\Theta_{j} + \Theta_{j+1}) \right] : \\
 \equiv - g \sum_{j} \int dx : \cos \left( 2 \vartheta_{j^{\ast}}  \right) :   \; ,
\label{eqn:inter-wire-int-Laughlin}
\end{multline}
where $\vartheta_{j^{\ast}}$ lives on the ``fictitious'' wire (or  strip) $j^{\ast}$ located in the middle of 
the two wires $j$ and $j+1$ (see the dashed lines in Fig.~\ref{fig:wire-vs-strip}).  
Physically, the part $\Phi_{j+1} - \Phi_{j}$ comes from the single-particle hopping between neighboring wires, 
and $m (\Theta_{j+1} + \Theta_{j})$ from the tailored backscattering {\em within} individual chains 
\cite{KaneMukhopadhyayLubensky,Teo-K-14}. 
The new fields $\{ \vartheta_{j^{\ast}} \}$ may be viewed as parts of a new set of bosons on strips (see Fig.~\ref{fig:wire-vs-strip})
\begin{equation}
\begin{split}
& 2 \varphi_{j^{\ast}} = \Phi_{j} + \Phi_{j+1} + m (\Theta_{j} - \Theta_{j+1})   \\
& 2 \vartheta_{j^{\ast}} = \Phi_{j} - \Phi_{j+1} + m (\Theta_{j} + \Theta_{j+1}) \; ,
\end{split}
\label{eqn:on-strinp-bosons}
\end{equation}
that, from \eqref{eqn:boson-commutation}, obey the following commutation relation:
\begin{equation}
[ \varphi_{j^{\ast}}(x) , \partial_x \vartheta_{j^{\prime\ast}}(x^{\prime}) ] 
= i\pi m\delta_{j^{\ast}j^{\prime\ast}}\delta(x - x^{\prime})  \; .
\label{eqn:CCR-strip-bosons}
\end{equation}
They may be viewed as made up of the new set of ``chiral bosons''
\begin{equation}
\tilde{\phi}^{\text{L}}_{j^{\ast}} = (\Phi_j + m \Theta_j)/2 , \;  
\tilde{\phi}^{\text{R}}_{j^{\ast}} = (\Phi_{j+1} - m \Theta_{j+1})/2
\label{eqn:modified-chiral-bosons-1}
\end{equation}
as $\varphi_{j^{\ast}} = \tilde{\phi}^{\text{L}}_{j^{\ast}}  + \tilde{\phi}^{\text{R}}_{j^{\ast}}$ 
and  $\vartheta_{j^{\ast}} = \tilde{\phi}^{\text{L}}_{j^{\ast}}  - \tilde{\phi}^{\text{R}}_{j^{\ast}}$.  

In the strong-coupling limit $g \to \infty$, the bosonic fields are pinned at one of the minima 
of the cosine potential, 
\begin{equation}
2 \bar{\vartheta}_{j^{\ast}}  
= 2 n_{j^{\ast}}  \pi  \quad (j^{\ast}=0,\dots, N_y -1) \; ,
\label{eqn:pinning}
\end{equation}
where $n_j \in \mathbb{Z} $.  Any semi-classical ground state is then specified by a set of $N_y$ integers $\{n_{j^{\ast}}\}$.  
However, all these states, labeled by $\{n_{j^{\ast}} \}$, are {\em not} physically distinct, since, as will be shown in the next section,  
some of them must be identified up to the periodicity of the bosonic fields $\Phi_j$ and $\Theta_j$.   
\subsection{Ground-state degeneracy}
\label{sec:GS-degeneracy}
One of the crucial signatures of a topologically ordered state in $2+1$ dimensions is the ground state degeneracy on higher-genus surfaces.  
The simplest non-trivial example is the $m$-fold degeneracy in the  $\nu = 1/m$ Laughlin states \cite{Haldane-R-85,Wen-N-90}.
In the strong-coupling limit, the (semi-classical) CWC ground states are found by minimizing the potential 
$- g \sum_{j} : \cos \left[ 2 \vartheta_{j^{\ast}}  \right]$ with respect to $\vartheta_{j^{\ast}}$, 
and there seems to be  infinitely many of them labeled by the set of integers $\{ n_{j^{\ast}} \}$ in \eqref{eqn:pinning}.  
 Reference~\cite{Teo-K-14}, gave an argument about the ground-state degeneracy based on the limit $N_y \to 1$.  
Here we count the number of inequivalent ground states by taking into account 
the periodic structure of the bosonic fields \cite{Lin-B-F-98,Lecheminant-T-06-SU4}.    
As the derivation is straightforward but slightly involved, we just give a sketch here and  
the details, for both fermions and bosons, are given in Appendix \ref{sec:couting-degeneracy}.  

First, we assume that the system is defined on a torus and impose periodic boundary condition both in the wire ($x$) direction 
and in the stacking direction ($N_y +1 \equiv 1$, $\vartheta_{0^{\ast}} \equiv \vartheta_{N^{\ast}_{y}}$).   
One may think that the wires are infinitely long and that the geometry of the system is cylindrical.  However, 
the argument (in Appendix \ref{sec:couting-degeneracy}) relies essentially on the structure of the zero modes which is peculiar to periodic systems 
(see, e.g., Fig.~\ref{fig:Gaussian_lattice}), and we implicitly assume the periodicity in the $x$ direction as well.  

Then we use the expressions \eqref{eqn:spinless-boson2fermion} expressing the Dirac fermions in terms of the bosons 
to infer that states which differ by a $\pi$ shift of $\phi_{j}^{\text{L,R}}$ must be identified:
\begin{equation}
\phi_{j}^{\text{L,R}} \simeq \phi_{j}^{\text{L,R}} + \pi \mathbb{Z}  \; .
\label{eqn:gauge-equivalence}
\end{equation}
(Bosons have a different periodicity, as explained in  Appendix \ref{sec:couting-degeneracy}.)
Due to the periodicity, most of the would-be  ground states are equivalent.  
In fact, as proven in Appendix \ref{sec:couting-degeneracy}, any pair of ground states $\{\bar{\vartheta}_{j^{\ast}}^{(1)}\}$ and $\{\bar{\vartheta}_{j^{\ast}}^{(2)}\}$
are equivalent and represent the same physical state if, and only if, 
\begin{equation}
\frac{1}{\pi}\sum_{j^{\ast}=1}^{N_y}  \delta \vartheta_{j^{\ast}} 
\equiv \frac{1}{\pi}\sum_{j^{\ast}=1}^{N_y } \left( \bar{\vartheta}^{(1)}_{j^{\ast}} -  \bar{\vartheta}^{(2)}_{j^{\ast}} \right)  =0  
\;\; (\text{mod } m) \, .
\label{eqn:gauge-redundancy-mod-m}
\end{equation}
Thus we conclude that in the coupled-wire system with strong inter-wire interaction \eqref{eqn:inter-wire-int-Laughlin},  
there are precisely $m$ distinct ground states that are characterized by:
\begin{equation}
\frac{1}{\pi}\sum_{j^{\ast}=1}^{N_y} \bar{\vartheta}_{j^{\ast}} =0,1,\ldots, m-1 \; .
\label{eqn:inequivalent-GS}
\end{equation}

From the above, it follows that the eigenvalues of the Wilson loop operator
\begin{equation}
\mathcal{W}_{y} = \exp\left[ i \frac{2}{m} \sum_{j^{\ast}=1}^{N_{y}} \vartheta_{j^{\ast}}(x_0) \right] \quad (x_0 \text{ : fixed}) 
\end{equation}
 distinguish the $m$ different ground states.   
Then, by using Eq.~\eqref{eqn:CCR-strip-bosons}, we readily see that the operator 
\begin{equation}
\mathcal{T}_{x} = \exp\left[ \frac{i}{m} \int_{0}^{L_x} \! dx \, \partial_{x}\varphi_{j_0^{\ast}}(x)  \right]  \quad (j^{\ast}_0 \text{ : fixed}) 
\end{equation}
changes the eigenvalue of $\mathcal{W}_{y} $ by $\be^{i \frac{2\pi}{m}}$ and creates a different ground state.   
Since, according to the arguments in Ref.~\cite{Teo-K-14}, the operator $\mathcal{T}_{x}$ may be viewed as transporting a quasi-particle along 
the wire $j^{\ast}_{0}$, this perfectly agrees with the well-known picture that the insertion of  fluxes, corresponding to 
quasi-particle transport around non-trivial loops, generates topologically different ground states \cite{Wen-N-90}.   
\section{From Coupled wires to Chern-Simons theory }
 \label{sec:gauge-from-CWC}

The argument in Sec.~\ref{sec:GS-degeneracy} already suggests us to identify $\vartheta_{j^{\ast}}$ with 
a gauge-like degree of freedom.  In this section, we proceed along this line to identify the bosonic fields with 
the statistical gauge field.
\subsection{Introducing gauge fields}
\label{sec:hidden-sym}
Since $x$-dependent local gauge transformations of fermions, $\psi_j \to \psi_j \be^{i \xi_j(x)}$, change the boson field as 
$\Phi_j (x) \to \Phi_j (x) + \xi_j(x)$, it is natural to identify 
\begin{equation}
a_ x (x,j l_y) = - \partial_x \Phi_j(x) \label{ax}
\end{equation}
as the $x$ component of a two dimensional gauge field. 
To obtain a bona fide two-dimensional gauge field, we also need the $y$ component, and to this end,   
we note that $\vartheta_j$ transforms as the $y$ component of a lattice gauge field.  
To see this, recall that a link variable transforms as \footnote{
Note that the $x$-dimension is continuous while the $y$-dimension is discrete. This can be thought of a an Euledian version of the Kogut-Susskind formulation of lattice gauge theory \cite{Kogut-S-75}. 
}
\begin{equation}
\begin{split}
U_{(x,j) , (x,j+1) } &\equiv e^{i l_y a_ y (x,j\delta) }   \\
& \rightarrow   e^{i\xi_j (x)} U_{(x,j) , (x,j+1)} e^{-i\xi_{j+1} (x) }  \; ,
\end{split}
\end{equation}
where the lattice spacing is the distance  $l_y$ between the wires. 
This implies
\begin{equation}
 a_ y (x,j l_y) \rightarrow a_ y (x,j l_y) + \frac{1}{l_y} (\xi_j (x) - \xi_{j+1}(x))  \, ,
\end{equation}
which suggests us to identify $2\vartheta_{j^\ast} (x) =  l_y a_y (x,j l_y)$ with a proper choice of gauge.  
In the continuum where we put $y=j l_y$, the $y$ component reads as:
\begin{equation}
a_y = - \partial_y \Phi + \frac {2m}{l_y} \Theta \, .   
\label{ay}
\end{equation}
Combining \eqref{ax} and \eqref{ay}, we obtain the following gauge field:
\begin{equation} \label{adef}
\vec a \equiv (a_x,a_y) = - \vec\nabla\Phi + (0, 2m \Theta /l_{y} ) \; , 
\end{equation}
which, by Eq.~\eqref{eqn:boson-commutation}, satisfies the commutation relation:
\begin{equation}
[a_x(\vec r_1), a_y(\vec r_2)] = i 2\pi m \delta^2(\vec r_1 - \vec r_2) 	\, .
\label{eqn:comm-rel-CS}
\end{equation}
This can also be derived from  a theory with a Chern-Simons term
\begin{equation}
{\mathcal L}_{\text{CS}} = \frac 1 {2\pi m} a_y \dot a_x \, , 
\label{eqn:CS-Lag}
\end{equation}
in the Lagrangian, which is precisely how the statistical gauge field appears 
in the Chern-Simons-Ginzburg-Landau (CSGL) theory \cite{Zhang-H-K-89}.
That the normalization of $\vec a$ is consistent with this interpretation is seen 
by calculating the statistical magnetic field
\begin{equation}
\frac{1}{2\pi} b_a = \frac{1}{2\pi} \epsilon^{ij} \partial_i a_j = \frac{m}{l_y} \delta \rho^{\text{1D}} = m \delta \rho  \, , 
\label{statb}
\end{equation}
where we have used Eq.~\eqref{eqn:1D-density-by-Theta} and that the two-dimensional density ($\rho$) is related to 
the one-dimensional one by $\rho^{\text{1D}} = l_y \rho$.  
Equation~\eqref{statb} implies that $m$ $a$ fluxes are attached to each electron thereby confirming our interpretation. 

An attentive reader should have noticed that in the above argument, the normalization of  $\vec a$ was of central importance. 
For instance, renormalizing the relation  \eqref{adef} by a factor $m$ would have given a coefficient $m/2\pi$ in the CS 
Lagrangian \eqref{eqn:CS-Lag}, and in that case it would have been tempting to identify $a$ with the hydrodynamical field in the Wen-Zee theory of quantum Hall liquids \cite{Wen-Z-92}.   
However, in the next subsection we show that our identification of $\vec a$ as a statistical gauge field is indeed the correct one.  

A remark on the strong-coupling limit is in order.  In the usual treatment of the CWC, the strong-coupling limit is 
used to simply gap out all the bulk degrees of freedom leaving the gapless modes only at the boundaries. 
Here, as we shall see, the strong-coupling limit enters via the consistency with the continuum limit.   In Ref.~\cite{Santos-H-G-G-15},  
an infinitely strong inter-wire interaction locks an \emph{external} electromagnetic field to $\vartheta$.   
For the external gauge field to be completely canceled by the statistical gauge field, which is the essence 
of the mean-field approximation in the CSGL theory,  the coupling must be strong. 
In the CSGL theory, the large energy scale is the cyclotron energy in the uncancelled magnetic field, while in our case it is the inter-wire coupling $g$.

\subsection{Relation to the CSGL theory}
\label{sec:CSGL}
So far, we have considered only the extreme strong-coupling limit $g \rightarrow \infty$ where $a_y$ 
is strictly pinned to the minimum of the cosine potential. Relaxing this condition, and also retaining the terms 
from the Luttinger Hamiltonian \eqref{eqn:Ham-Dirac}, we get the effective Hamiltonian for the $a$ field
\begin{equation}
H_{\text{eff}} 
= \int d^2 r\,  \left\{ \frac {\omega_{\text{c}}} {4\pi m}  \left( \frac 1 K a_x^2 +   2 m \beta a_y^2\right) 
+ \frac{K \omega_{\text{c}} \pi}{4m}l_y^2 (\delta\rho)^2  \right\}  \; ,
\label{effaham}
\end{equation}
where we used the mean-field relation $2\pi m \bar\rho = e B$, with  $B$  the external magnetic field and $\bar\rho$ the mean {\em area} density,  
to rewrite the Fermi velocity as 
$v_{\text{F}} = \pi \bar{\rho}^{\text{1D}}/ M_{\text{e}} = \omega_{\text{c}}l_y \nu/2=\omega_{\text{c}} l_y /(2 m)$. 
Here, $\omega_{\text{c}}=eB/M_{\text{e}}$ is the cyclotron energy ($M_{\text{e}}$ is the electron band mass), 
and we introduced the dimensionless coupling strength $\beta$ by,
\begin{equation}
g = \frac{\omega_{\text{c}}}{\pi l_y} \beta \, .
\end{equation}
In the last term in \eqref{effaham}, which comes from the term $\sim (\partial_x\Theta)^2$ in the Luttinger Hamiltonian \eqref{eqn:Ham-Dirac}, 
$\delta\rho$ denotes the deviation from the average area density $\bar\rho = e B /(2\pi m)$.\\

So far, $K$ and $\beta$ are free parameters but we shall now impose the condition $\beta K = 1/(2m)$ 
which ensures  rotational invariance in \eqref{effaham}. It should be no surprise that fine tuning is needed to retain  rotational invariance;
it is in fact more surprising that rotational invariance can at all be recovered in the CWC. 
In order for the continuum approximation to make sense [for $\beta=1/(2mK)=\text{finite}$], 
we must take $l_y$ to be small and thus $g$ must be large $g \sim 1/l_y$,  albeit not infinite. 
The rotationally invariant effective Lagrangian now reads as
\begin{equation} 
{\cal L}_{\text{CWC}} =  \frac 1 {4\pi m} \epsilon^{ij}a_i \dot a_j  - \frac{\beta}{2\pi} \omega_{\text{c}} ( a_x^2 + a_y^2) 
- \frac{K \omega_{\text{c}} \pi}{4m}l_y^2 (\delta\rho)^2  \, . 
 \label{effalag}
\end{equation}
This is \emph{not} a topological field theory since the leading term ($a_i^2$)  in a derivative expansion does depend on the metric.  
Thus we could not have interpreted $a$ as the hydrodynamical field even if we had changed the normalization 
to get the ``correct'' CS term. 

It is illuminating to compare this with the standard CSGL theory given by the Lagrangian \cite{Zhang-H-K-89,Zhang-CSGL-92}, 
\begin{equation}
\begin{split}
&{\cal L}_{\text{CSGL}} = \phi^{\ast}(i\partial_{0}- \alpha_{0})\phi 
- \frac{1}{2M_{\text{e}}}\bigl| ( - i \vec\nabla - e\vec{A}-\vec{\alpha} )\phi \bigr|^{2}  \\
&  -   \frac{1}{2} \int \! d \vec{y} \,  |\phi(\vec{x})|^{2} V(\vec{x}-\vec{y}) |\phi(\vec{y})|^{2}  + \frac 1 {4\pi m} \epsilon^{\mu\nu\lambda} \alpha_\mu\partial_\nu \alpha_\lambda \, , 
\label{philag}
\end{split}
\end{equation}
where $\phi$ is a composite boson minimally coupled to the external electromagnetic gauge field $\vec A$,  $(\alpha_0,\vec\alpha)$ is a statistical gauge field,  and $V(\vec r)$ is a repulsive two-body repulsive potential.

We proceed by first parametrizing the bosonic field $\phi$ as 
\begin{align}
\phi (\vec{x}) =\sqrt{\rho(\vec{x})}e^{i\theta(\vec{x})}    \, ,
\label{phipar}
\end{align}
and then integrate out the non-dynamical field $\alpha_0$ to get  the constraint $\rho = 2\pi m b_{\alpha}$, where $b_{\alpha}=\epsilon^{ij}\partial_i \alpha_j$ 
is the flux of the statistical gauge field $\vec\alpha$, and the chemical potential is such that $V$ has a minimum at $\rho = \bar\rho$.
Neglecting derivatives of the density,  making the mean-field approximation $\vec a = e\vec A + \vec \alpha $, and assuming a 
local potential $V(\vec r) =V \delta(\vec r)$, we get 
\begin{equation}
{\mathcal L}_{\text{CSGL}} =  \frac 1 {4\pi m} \epsilon^{ij} \alpha_i \dot \alpha_j 
- \frac {\bar\rho} {2M_{\text{e}}} \vec \alpha \cdot  \vec \alpha - \frac{V}{2} (\delta\rho)^2   
\label{csgllag}
\end{equation}
with $2\pi m \bar\rho = eB$. To derive this result, we absorbed the gradient of the phase, $\vec\nabla\theta$,  
in the vector potential $\vec \alpha$, which amounts to picking a unitary gauge. 

Recalling that $\bar \rho/M_{\text{e}} = \omega_{\text{c}}/(2\pi m)$, \eqref{effalag} and \eqref{csgllag} 
become identical if we take $\beta =1/(2m)$ (i.e., $K=1$), identify  $\vec a$  in \eqref{effalag} as the fluctuation 
in the statistical gauge field, and pick $V$ in the CSGL theory \eqref{csgllag}
 to match the coefficient in front of $(\delta\rho)^2$ in the CWC expression.
In what follows, we fix $K=1$ and $\beta = 1/(2m)$ except in Sec.~\ref{sec:general-TQO-2D}.
This equivalence is a main result of this paper, and it shows that, after introducing interactions,  
the bosons on the wires become bona fide composite bosons.   

Having shown the equivalence between the CWC and  the CSGL theory, it is reasonable to expect that the results from the latter can be derived also in the present context.  In the next two sections, we shall show that this is indeed the case. 
Before doing so, however, we shall make some comment about the edge modes.

The presence of  chiral gapless edge modes is at the heart of the CWC.  
The way the right and left moving modes on the adjacent wires are coupled is designed in such a way 
that the correct chiral modes are left at the wires at the edges of the system.  
It is clearly of importance that our continuum description in terms of gauge fields is able to describe these modes as well.   
In fact, as we already advertised, we will eventually derive the Wen-Zee theory, which is known to give 
a correct description of the edge \cite{Wen-95}.  
Turning to the intermediate theory ${\cal L}_{\text{CWC}}$ given by \eqref{effalag}, 
one might naively expect that there are no gapless modes because of the quadratic terms. 

\section{ The Laughlin wave function from coupled wires  }  
\label{sec:Laughlin-wf}
 In this section, we derive the Laughlin wave function from the CWC.  
 We do it first for the (fine-tuned) rotationally invariant theory \eqref{effalag}, and then for the general anisotropic case.  
\subsection{The isotropic case}
\label{sec:Laughlin-wf-isotropic}
In the standard derivation \cite{Kane-K-L-Z-91,Zhang-CSGL-92}  
of the wave function from the CSLG theory, one uses the Coulomb gauge 
and rewrites the Hamiltonian as a collection of harmonic oscillators in the variables $\delta\rho$ and $\theta$.  
One then finds the wave function for the composite bosons, by using the density representation of the wave functional.  
Finally, the full lowest-Landau-level Laughlin wave function is regained by reintroducing the phase factor 
that was taken out in the statistical transmutation from the original electrons to the composite bosons.  
In the present context, the derivation of the norm of the wave functions only differs from the standard CSGL 
procedure  by some  technicalities, while that for the phase requires a careful treatment,  
and provides a non-trivial consistency check of our calculations.

Instead of the Coulomb gauge, we are implicitly using the unitary gauge, where the phase fluctuations are absorbed 
in the gauge field [see, e.g., Eq.~\eqref{csgllag}], and our treatment here also differs from that in the standard CSLG treatment in that the relation 
between $\vec a$ and the density is anisotropic as is seen in Eq.~\eqref{statb}. 
With this in mind, we proceed by making the following decomposition for $\vec{a}$: 
\begin{equation} \label{adecomp}
 a_i   = \epsilon_{ij} \partial_j \chi + \partial_i \eta
 \end{equation}
 so that $b_a =-\nabla^2 \chi $.
Substituting \eqref{adecomp} in the Lagrangian density \eqref{effalag} gives  the Lagrangian,
\begin{equation}
\begin{split}
L_{\text{CWC}} =&  \frac 1 {2\pi m}  \int d^2 r\,  \eta\dot b_a
-   \frac {\omega_{\text{c}}} {4\pi m}\int d^2r\, \left( -\eta\nabla^2\eta - \chi \nabla^2 \chi \right)   \\
&- \frac{ \omega_{\text{c}} \pi}{4m}l_y^2   \int d^2 r\,  (\delta\rho)^2  \, , 
\end{split}
 \label{lagrangian}
\end{equation}
where we have set $\beta = 1/ (2m)$ and used that, by partial integration, the cross terms between $\eta$ and $\chi$ vanish in the Hamiltonian part  
while the diagonal terms vanish in the kinetic term.
Using $b_a =  2\pi m \delta \rho$, we see that $\delta\rho$ and $\eta$ are conjugate variables 
satisfying $ [\delta \rho(\vec r_1), \eta(\vec r_2)] =  i \delta^2(\vec r_1 - \vec r_2)$. The Hamiltonian becomes,
\begin{equation}
 H=  \frac {\omega_{\text{c}}} {4\pi m}\int d^2r\, \left\{  -\eta\nabla^2\eta - \delta \rho \frac {(2\pi m)^2} {\nabla^2} \delta \rho \right\}   \, ,
\label{wfham}
\end{equation}
where we  used $\beta=1/(2m)$ and only kept the leading terms in a derivative expansion. 
In the density ($\delta \rho$) representation, the wave functional is that of an assembly  of harmonic oscillators labeled by $r$.  
Using the explicit non-relativistic form of the density operator, 
$\delta \rho(\vec r) = \rho(\vec r) - \bar\rho = \sum_i \delta^2(\vec r - \vec r_i) - \bar\rho$ with $\vec r_i $ being 
the position of the $i^{\text{th}}$ electron, and the mean density $\bar\rho$ which is related to the magnetic length $\ell_{\text{B}}$ 
by $2\pi m \bar\rho = eB = 1/\ell_{\text{B}}^2$,  the wave function reads as
\begin{equation} \label{mlwf}
 \Psi_m  (\{ \vec{r}_{i} \}) \sim \prod_{i<j} | z_i - z_j|^m e^{-\frac 1 {4 \ell_{\text{B}}^2}\sum_i |z_i|^2}\, .  
 \end{equation}
In the above, we have used the complex notaion $z_i = x_i + iy_i$, 
and we refer to Refs.~\cite{Kane-K-L-Z-91,Zhang-CSGL-92}  for details of the derivation. 

\begin{figure}
\begin{centering}
\includegraphics[width=0.8\linewidth]{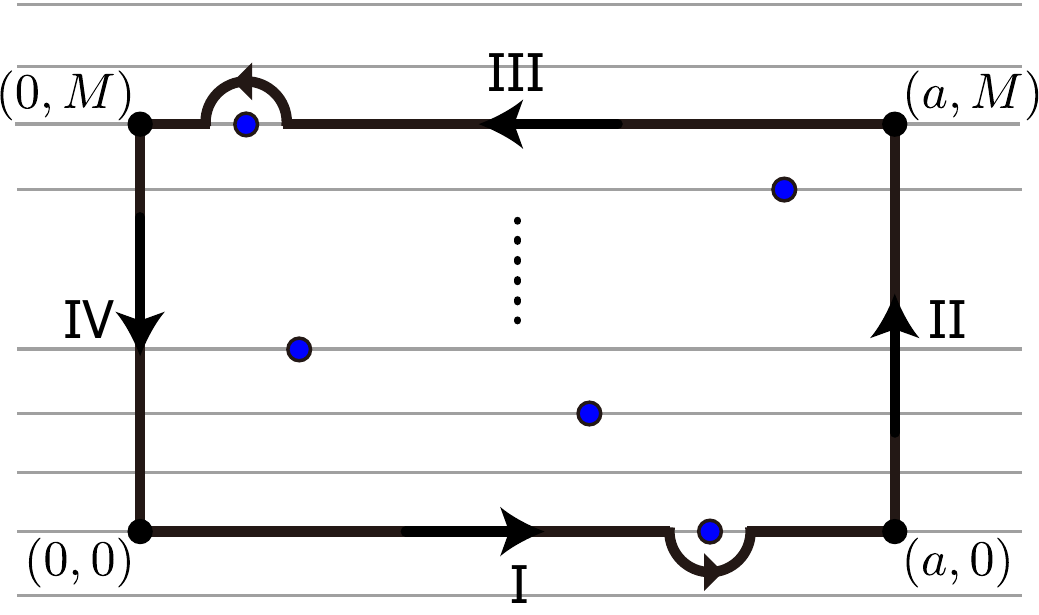}
\par\end{centering}
\caption{ %
(Color online) Loop used to calculate the phase of the Laughlin wave function.  
Particles on wires 1 to $M-1$ are fully encircled and each gives the phase $2m\pi$.  
Exchanging the positions with particles  on wires 1 or $M$ amounts to replacing the straight lines with half-circles, 
each one giving rise to a phase $m\pi$. 
\label{fig:loop}}
\end{figure}

The Laughlin wave function differs from \eqref{mlwf} by the phase factor 
\begin{equation} \label{ephasefactor}
\prod_{i<j} (z_i - z_j)^{\frac m 2} / \prod_{i<j} (\bar z_i - \bar z_j)^{-\frac m 2} \, ,
\end{equation}
in which exchanging two electrons yields a phase angle $m\pi$, while transporting one around another yields $2m\pi$.   
We now show that we can reconstruct the additional phase factor \eqref{ephasefactor} within the CWC by carefully calculating 
the Berry phase factor acquired  when transporting an electron along the path shown in Fig.~\ref{fig:loop}.   
In doing so, we first calculate the phase using Eq.~\eqref{ephasefactor}.    
When moving along the paths I and III, the transported electron exchanges its position with other electrons sitting on the wires 0 and $M$ 
thereby acquiring the phase $m \pi$ each time (in the $\nu = 1/m$ Laughlin state).   
On the segments II and IV, on the other hand, there are no exchanges.  Nevertheless, the full loop encircles all the electrons sitting 
on the wires 1 to $M-1$ and picks up the phase $2m \pi$ from each electron.  
If we define $\theta_j(x)$ to denote the ground-state expectation value of the operator $\Theta_j(x)$, 
then, by Eq.~\eqref{eqn:1D-density-by-Theta}, the number of electrons in the interval 
$x \in [0,a]$ of the wire $j$ is given 
by $(\theta_j(a) - \theta_j(0))/\pi$.  
Combining all this, we expect the total phase acquired by the electron: 
\begin{equation} 
\begin{split}
\gamma_{\text{loop}} =& m \{ \theta_0(a) - \theta_0 (0) + \theta_M(0) - \theta_M (a) \}   \\ 
&+ 2m \sum_{j=1}^{M-1} \{ \theta_j(a) - \theta_j (0) \}  \, .
\end{split}
\label{loopres}
\end{equation}

To calculate the Berry phase related to transporting a right-moving electron (we could equally well have considered a left mover) 
within the CWC, we first define the states:
\begin{equation}
\ket {\vec r} = \be^{i \Phi_j (x) - i \Theta_j (x) } \ket 0 \equiv \be^{i\varphi(\vec r)} \ket 0
\end{equation}
 where $\vec r = (x,j  )$ and $\ket 0$ is the ground state.   We have also introduced the notation 
 $\varphi (\vec r) = 2\phi^{\text{R}}_{j}(x)$ to avoid clutter.   
 The Berry connection is defined by:
 \begin{equation}
 \begin{split}
 \vec {\mathcal A} (\vec r) & =  \bra {\vec r} i \nabla_{\vec r} \ket{\vec r} = \bra 0 \be^{-i\varphi (\vec r)}  i \nabla_{\vec r} \, \be^{i\varphi(\vec r)} \ket 0  \\
 & = \bra 0 \partial_{x} \Theta_{j}(x) \ket 0\, ,
 \end{split}
 \end{equation}
 where we  used that, because of reflection symmetry when $x\rightarrow -x$, there is no $x$-component of the current in the ground state: 
$\bra{0} \partial_x \Phi_j (x) \ket 0 = 0$, and properly regularized the product $\be^{-i\varphi (\vec r)}\be^{i\varphi(\vec r)}$.
It is now easy to calculate the Berry phase corresponding to  the segment I:
\begin{equation}
\Upsilon(\text{I}) = \int_0^a  dx \, {\mathcal A}_x (x,0) 
= - ( \theta_0(0) - \theta_0(a) )  \; .
\end{equation}
In the same way, we get for the segment III:
\begin{equation}
\Upsilon (\text{III})  =  \theta_M (0) - \theta_M (a)  \; .
\end{equation}

Along to the segment II, we shall return to the original discretized formulation and define the the Berry phase factor as we do in lattice field theories
\begin{eqnarray} 
\be^{i \Upsilon (\text{II}) } =&  \braket { (a,M)} {(a, M-1)} \braket { (a,M-1)} {(a, M-2)}  \nonumber \\
&  \phantom{=} \quad  \dots \braket { (a,1)} {(a, 0)}  
\end{eqnarray}
with
\begin{equation}
\braket {(x,j+1)} {(x,j)} = \bra 0 \be^{-i(\varphi(x,j+1) - \varphi(x,j))}\ket 0
 \label{vertel}
\end{equation}
where we used that $\varphi(x,j)$ on different wires commute.  
 The next, and crucial, step is to rewrite the exponent in \eqref{vertel} as,
 \begin{equation}
 \begin{split}
 & \varphi(j+1) - \varphi(j) \\
 & = \Phi_{j+1} - \Phi_j - m(\Theta_{j+1} + \Theta_j)  \\
& \phantom{=} \qquad \qquad \qquad 
 + (m -1) \Theta_{j+1} + (m+1) \Theta_j \\
 & = - 2\vartheta_{j^{\ast}}  + (m + 1) \Theta_j + (m - 1) \Theta_{j+1}   \; ,
 \end{split}
 \end{equation}
 where we have suppressed the $x$ dependence. We now recall, from Sec.~\ref{IIA}, that $\vartheta_{j^{\ast}}$ 
 is pinned to a constant $\bar{\vartheta}_{j^{\ast}}$,  
 which we take to be zero (as $2\bar{\vartheta}_{j^{\ast}} \equiv 0$ mod $2\pi$, we arrive at the same phase factor for any other choices; 
 the important thing is that it does not depend  
 on the geometry of the loop). Adding the contributions from all the wires ($j=0,\dots,M-1$) together 
 and taking the expectation value, we are left with the phase:
 \begin{equation}
 \begin{split}
\Upsilon (\text{II}) = & - \bigl\{ (m + 1)\theta_0 (a) + 2m [ \theta_1(a)  + \\
& \qquad \dots + \theta_{M-1}(a) ] + (m - 1) \theta_{M} (a)  \bigr\} \, .
\end{split}
\end{equation}
The segment IV gives a similar expression, now evaluated at $x=0$, and with an additional minus sign.  
 Putting everything together, and taking into account that the Berry phase differs by a sign from the 
 exchange phase calculated from the wave function \cite{Hansson-S-L-90},
 we obtain precisely the phase \eqref{loopres} that was derived from \eqref{mlwf}.  
 Note that, only after carefully taking into account both the phases $m\pi$ from the particle exchange {\em on} the loop and $2m \pi$  
 from the encircled charges {\em inside}, we obtained the correct result.  

 
 Because of the close analogy to the CSGL theory, it is natural to expect that the above derivation can also be modified 
 to give the wave functions of the Laughlin holes. We shall not elaborate on the details, but just make two observations.  
First, adding a kink $\Theta_{j_\eta}^{\text{kink}}(x) = \pi \, \mathrm{sgn} (x - \eta_x)/(2m)$ 
 on the wire $j_{\eta}$ in the loop in Fig. \ref{fig:loop}, will, by exactly the same argument as above, give 
the Berry phase \eqref{loopres} without $m$ thereby leading to an extra phase factor:
 $$\prod_{i} (z_i - \eta)^{\frac 1 2} / \prod_{i} (\bar z_i - \bar \eta)^{-\frac 1 2}  \, ,$$
 where $\eta = \eta_x + i j_{\eta} l_y$. Since it will also give a contribution $\delta^{2}(\vec r - \vec \eta)/m$ [$\vec{\eta}=(\eta_x, j_{\eta}l_y)$] 
 to the density operator $\rho(\vec r)$, it is easy to see that it will add the correct modulus $\prod_{i} | z_i - \eta|$ 
 which, together with the above phase factor, reproduces the correct Laughlin hole factor $\prod_{i} (z_i - \eta)$.
 
 The quasielectrons are harder to deal with. Although it is straightforward to calculate the fractional charges 
 and the statistical phase factors for states with anti-kinks on the wires, it is non-trivial to extract  wave functions.  
 This asymmetry in the description of quasiholes and quasielectrons is well-known both in the CSGL theory, and in 
 approaches based on conformal field theory \cite{Hansson-H-S-V-17}.

\subsection{The anisotropic case}
If we do not make  the (fine-tuned) choice $\beta K = 1/(2m)$, 
that is necessary to get the isotropic theory \eqref{effalag}, the Hamiltonian will depend on the combination 
$$
a_x^2 + \alpha^2 a_y^2
$$
with $\alpha^2 = 2m \beta K$.  
Since $a_x = -\partial_x \Phi$ and $a_y  = -\partial_y \Phi + \frac {2m} \ell \Theta$, a rescaling $(x,y) \rightarrow (x',y') = (\alpha x,y)$ 
implies,
\begin{equation}
\begin{split}
& a_x(x,y) \rightarrow  a^{\prime}_x (x^{\prime},y^{\prime}) = \alpha^{-1} a_x(x ,y) \\
& a_y(x,y) \rightarrow   a^{\prime}_y (x^{\prime},y^{\prime}) = a_y(x ,y) \, .
\end{split}
\end{equation}
Alternatively, we could rescale $(x,y) \rightarrow (x',y') = (x, y/\alpha )$  and $l_y \to l^{\prime}_{y} = l_y/\alpha$. In both cases, 
the $a^{2}$ terms change as:
\begin{equation}
dx\,dy\, (a_x^2 + \alpha^2 a_y^2) \rightarrow \alpha dx^{\prime}\,dy^{\prime} \, \left\{ (a^{\prime}_x)^2 + (a^{\prime}_y)^2 \right\}  \; ,
\end{equation}
while $dx\,dy\, a_x \dot a_y$ remains invariant. Thus, if we use the primed coordinates, the previous derivation of the Laughlin wave function 
will go through without any changes, while the complex coordinate will be redefined as $z = x + i y/\alpha$.  
In terms of the wave functions in the lowest Landau level, this amounts to having coherent states which are not circular,  
but deformed into ellipses.  
As has been stressed by Haldane \cite{Haldane2011} and others \cite{suorsa2011,Hansson-H-S-V-17}, the Laughlin state is more general in nature than 
its usual incarnation as a holomorphic polynomial $\prod_{i<j}(z_i-z_j)^m$ in $z=x+iy$. 

\section{The topological and hydrodynamic actions}
\label{sec:hydrodynamic-WZ}
 Here we derive the effective hydrodynamical theory which contains the Wen-Zee topological field theory 
 as the leading term in the infrared.  From this, we  extract the collective Kohn mode and its dispersion.  
We also show how to couple the hydrodynamic theory to an external perturbing electromagnetic field. 

\subsection{The topological action}
\label{sec:topological-action}
When doing the gauge transformation 
\begin{equation} 
 a_i \rightarrow  a_i + \partial_i \xi \, , \label{gtransf}
 \end{equation}
the CS Lagrangian \eqref{eqn:CS-Lag} picks up an extra piece $\sim \epsilon^{ij}a_i\partial_j \dot \xi$,  
and, using the notation $a_0 = \dot\xi$, we can recast it into the relativistic form,
\begin{equation}
{\mathcal L}_{\text{CS}} (a) = \frac 1 {4\pi m} \epsilon^{\mu\nu\lambda} a_\mu\partial_\nu a_\lambda \, .   
\label{eqn:CS-Lag-relativistic}
\end{equation}
If we introduce a new gauge field $b$, the partition function for this CS theory can be rewritten as follows,
\begin{equation}
Z = \int {\mathcal D}[a] e^{i S_{\text{CS}}[a]} =  \int {\mathcal D} [a] {\mathcal D} [b] e^{i S_{\text{eff}}[a,b]}  \; ,
\end{equation}
where $S_{\text{CS}}[a] = \int d^3r \, {\mathcal L}_{\text{CS}}(a) $, and 
$S_{\text{eff}}[a,b] = \int d^3 r\, {\mathcal L}_{\text{eff}}(a,b)$ with 
\begin{equation}
 {\mathcal L}_{\text{eff}}(a,b) = - \frac m {4\pi } \epsilon^{\mu\nu\lambda} b_\mu\partial_\nu b_\lambda  
 + \frac 1 {2\pi} \epsilon^{\mu\nu\lambda} a_\mu\partial_\nu b_\lambda \, .  
 \label{eqn:WZ-action} 
\end{equation}
Integrating out the statistical gauge field $a$ in \eqref{eqn:WZ-action} yields the constraint $\epsilon^{\mu\nu\lambda} \partial_\nu b_\lambda =0$ 
(remember that the background charge has been subtracted already) leaving only the first term  
which is precisely the Wen-Zee topological action for the $\nu = 1/m$ Laughlin state.  

The Wen-Zee action is trivial on an infinite plane, but codes for the $m^{g}$-fold ground-state 
degeneracy on higher-genus ($g$) surfaces, as well as giving the kinetic term for the chiral edge modes \cite{Wen-95}.  
The ground state degeneracy is consistent with the counting done in  Sec.~\ref{sec:GS-degeneracy}, and the presence of chiral 
edge modes in the direction of the wires was a starting point of the CWC \cite{KaneMukhopadhyayLubensky}.  

The fact that the Wen-Zee action follows from the CWC 
strongly suggests that there will be gapless edge modes also in the perpendicular 
direction in the case of finite length wires with open boundary conditions. 
However, we do not claim to have proven this since the derivation of the topological action 
did not incorporate open boundary conditions in the $x$-direction.   
Here, a comment on the edge Hamiltonian is in order. From the CWC perspective, the edge modes parallel 
to the wires are nothing but the decoupled chiral components at the outermost wires.  
As such, they have a Hamiltonian given explicitly by the chiral Luttinger theory which is at the basis of the construction.  
In particular, the velocity depends explicitly on the topological number $m$. There is no reason to believe 
that this would give a good description of a {\em real} QH edge, where the edge velocity is known to depend 
on the edge potential. Thus, the edge velocity should be considered as an extra phenomenological parameter, 
and the same holds for the multi-component case discussed in Sec.~\ref{sec:general-TQO-2D}.

\subsection{The hydrodynamical action and the Kohn mode}
\label{sec:Kohn-mode}
In the previous section, we only retained the topological part \eqref{eqn:CS-Lag} of the full action \eqref{effalag}.   
We now show that, by including the Hamiltonian part, we can complement the Wen-Zee topological term with higher derivative contributions  
that describe collective modes. To this end, we begin with the low-energy effective Lagrangian \eqref{effalag}:  
\begin{equation}
\begin{split} 
\mathcal{L}_{\text{CWC}} = & \frac 1 {4\pi m} \epsilon^{\mu\nu\lambda} a_\mu\partial_\nu a_\lambda -a_0\delta\rho   \\
&- \frac \beta {2\pi} \omega_{\text{c}} ( a_x^2 + a_y^2) 
- \frac{ \omega_{\text{c}} \pi}{4m}l_y^2 (\delta\rho)^2  \, ,  
\label{effalag2}
\end{split}
\end{equation}
where we have added the term $a_0\delta \rho$ to retain the correct constraint. 
Following the same logic as in the previous section, but with ${\mathcal L}_{\text{CS}}(a)$ 
replaced by the above expression, we first integrate out the density fluctuations $\delta \rho$ to get a term $\sim a_0^2$ in the action: 
\begin{multline} 
{\mathcal L} = - \frac m {4\pi} \epsilon^{\mu\nu\lambda} b_\mu\partial_\nu b_\lambda + \frac 1 {2\pi} \epsilon^{\mu\nu\lambda} 
a_\mu\partial_\nu b_\lambda \\ - \frac {\omega_{\text{c}}} {4\pi m} \vec a \cdot \vec a  + \frac {m\omega_{\text{c}}} {\pi\gamma} a_0^2  \; ,
\label{effalag3}
\end{multline}
where we  defined the dimensionless parameter {$\gamma = K \omega_{\text{c}}^{2} l_y^2$ and used the relation $\beta=1/(2m)$.   

As above, we then proceed to integrate the $a$ field to get the desired effective hydrodynamic theory:
\begin{multline} 
{\cal L}_{\text{hydro}}  = - \frac m {4\pi} \epsilon^{\mu\nu\lambda} b_\mu\partial_\nu b_\lambda 
+ \frac m {4\pi} \frac{1}{\omega_{\text{c}}} \vec E_b^2 - \frac {\gamma} {16\pi m} \frac 1 {\omega_{\text{c}}} B_b^2  \, ,
\label{hydro}
\end{multline} 
where $\vec E_b = -\vec\nabla b_0 - \dot{\vec{b}}$ and $B_b = \epsilon_{ij}\partial_i b_j $ are the field strengths related to $b$. 

It is now straightforward to extract the dispersion relation for the collective mode described by the dynamics of this effective theory,
\begin{equation}
\omega (q) = \omega_{\text{c}} + \frac {\gamma} {8m^2 } \frac 1 {\omega_{\text{c}}} q^2
\end{equation}
This should be compared with the result from the CSGL theory \eqref{philag} \cite{Lee-Z-91,Zhang-CSGL-92}:
\begin{equation}
\omega (q) = \omega_{\text{c}} + \frac {\bar\rho} {2B} V(q)q^2  \; ,
\end{equation}
where $V(q) $ is the Fourier transform of the two-body potential $V(\vec r)$.  
We see, as  expected from \eqref{effalag} and \eqref{csgllag}, 
that the result from the coupled wire construction correspond to having a delta function potential.  

\subsection{Coupling to an external electromagnetic field}
So far we did not include the coupling to electromagnetism, except for the constant $B$ field 
that determines the ground state density $\bar{\rho}$.  
Minimal coupling of electromagnetism to a phase field is implemented by the standard substitution,
\begin{equation} \label{substitution}
\partial_i \Phi \rightarrow \partial_i \Phi -eA_i  \, ,
\end{equation} 
and recalling the definition \eqref{adef}, this implies that $\vec A$ is incorporated by the substitution $\vec a \rightarrow \vec a + e\vec A$ 
in the Hamiltonian, and $A_0$ is coupled by adding the term $eA_0 \rho$ to \eqref{effalag2}.  
With this in mind, the Lagrangian \eqref{effalag3} generalizes to,
\begin{equation}
\begin{split} 
{\mathcal L} =&  - \frac m {4\pi} \epsilon^{\mu\nu\lambda} b_\mu\partial_\nu b_\lambda 
+ \frac 1 {2\pi} \epsilon^{\mu\nu\lambda} a_\mu\partial_\nu b_\lambda \\ 
& - \frac {\omega_{\text{c}}} {4\pi m} (\vec a + e\vec A)\cdot (\vec a + e\vec A) + \frac {m\omega_{\text{c}}} {\pi\gamma} (a_0 + e A_0)^2 \, .
\end{split}
\end{equation}
 [Note that the substitution \eqref{substitution}, which amounts to a minimal coupling, should be done {\em only} in the Hamiltonian;   
 the CS action, which encodes the proper commutation relations, is not to be changed.]  
By a shift, $a\rightarrow \tilde a - eA$, the integration over $\tilde a$ can be performed as in the previous section,  
and we regain the result  \eqref{hydro} with the extra term
\begin{equation}
{\mathcal L}_A = - \frac e {2\pi} \epsilon^{\mu\nu\lambda} A_\mu\partial_\nu b_\lambda \, ,  \label{emcoupl}
\end{equation}
which is the desired coupling of the electromagnetic potential to the conserved current 
$j^\mu =  \frac e {2\pi} \epsilon^{\mu\nu\lambda} \partial_\nu b_\lambda$.    

At this point, it behooves us to clarify the relation to the approach  by Fuji and Furusaki in Ref. \cite{Fuji-F-19}.  
Our hydrodynamic Lagrangian \eqref{hydro}, supplemented by the electromagnetic coupling \eqref{emcoupl}, should be 
compared with their Eq.~(52).  
They use the notation $\alpha$ for our hydrodynamic gauge field $b$ and also consider couplings to 
a bosonic quasiparticle described by $\Phi^{\text{VCB}}$.  
As they point out, their final expression contains a term that may be obtained by discretizing the topological part of \eqref{hydro}. 
However, it crucially differs from \eqref{hydro} in that the field $\alpha$ does not have any dynamics, 
and that Eq.~(52) has other terms that may eventually generate more relevant contributions (e.g., $\alpha^{2}$) in a derivative expansion.

\section{General abelian QH states}
\label{sec:general-TQO-2D}
Having understood how a topological quantum field theory emerges from the CWC in the simplest case of the Laughlin states,
we now ask what in the previous analysis will carry over to general abelian QH states which are described by 
the Wen-Zee Lagrangian \cite{Wen-Z-92}:
\begin{equation}
\mathcal{L} = - \frac{1}{4\pi} \sum_{I,J=1}^{n_K} \mathcal{K}_{IJ} \epsilon^{\mu\nu\rho} b^{(I)}_{\mu}\partial_{\nu} b^{(J)}_{\rho} 
- \frac{e}{2\pi} \sum_{I=1}^{n} q_{I} \epsilon^{\mu\nu\rho} b^{(I)}_{\mu}\partial_{\nu} A_{\rho}    \; .
\label{eqn:def-multi-comp-CS}
\end{equation}
In the above, $\{ b^{(I)}_{\mu} \}$ are $n_K$  Chern-Simons gauge fields,  and the $n_K{\times}n_K$ ``$K$ matrix'', $\mathcal{K}$,  
is symmetric and integer-valued.  
The integer-valued ``charge vector'', $\mathbf{q}=(q_1,\ldots,q_{n})$ determines how the $I$th component of the U(1) charge 
$\frac{1}{2\pi}\epsilon^{\mu\nu\rho} \partial_{\mu}b^{(I)}_{\nu}$ couples to the external elcetromagnetic field $A_{\mu}$ \cite{Wen-95}.   
The topological field theories of the form \eqref{eqn:def-multi-comp-CS} are known to describe not only genuine topological 
phases \cite{Wen-Z-92} but also symmetry-protected ones \cite{Lu-V-12}.  
In Ref.~\cite{Teo-K-14}, Teo and Kane gave a construction for the simple case of a $2{\times}2$ $K$ matrix.    
Their method generalizes quite straightforwardly to a general $K$ matrix, and so does the extraction of the topological field theory 
\eqref{eqn:def-multi-comp-CS}.   

To closely follow the steps given in Sec.~\ref{IIA} for the single component case, we prepare
 $N_{\text{layer}}=n_K$ layers of coupled wires and define the following bosons as in Eq.~\eqref{eqn:modified-chiral-bosons-1}, 
\begin{equation}
\vec{\widetilde{\phi}}{}^{\text{L}}_{j^{\ast}} \equiv \frac{1}{2} \left( \vec{\Phi}_{j} + \mathcal{K} \vec{\Theta}_{j} \right)  \; , \;\;
\vec{\widetilde{\phi}}{}^{\text{R}}_{j^{\ast}} \equiv \frac{1}{2} \left( \vec{\Phi}_{j+1} - \mathcal{K} \vec{\Theta}_{j+1} \right) \; ,
\label{eqn:modified-chiral-multi-comp}
\end{equation}
where we have introduced the vectorial notations, e.g., 
\begin{equation}
 \vec{\Phi}_{j} \equiv  \left(\Phi_{j}^{(1)},\ldots, \Phi_{j}^{(N_{\text{layer}})} \right)^{\text{T}} \; , \;\; 
 \vec{\Theta}_{j} \equiv  \left(\Theta_{j}^{(1)},\ldots, \Theta_{j}^{(N_{\text{layer}})} \right)^{\text{T}}   \; .
 \end{equation}
Each pair of bosons $( \Phi^{(I)}_{j},\Theta^{(I)}_{j} )$ obey the commutation relations \eqref{eqn:boson-commutation} and 
describe a Luttinger liquid on the wire-$j$ of the $I$th layer (see Fig.~\ref{fig:CWC-double-layer2}) 
with the Luttinger parameter $K$.
From these fields, we define the following bosons living on the strip $j^{\ast}$:
 \begin{equation}
 \begin{split}
\vec{ \vartheta}_{j^{\ast}} \equiv \vec{\widetilde{\phi}}{}^{\text{L}}_{j^{\ast}}  -  \vec{\widetilde{\phi}}{}^{\text{R}}_{j^{\ast}}   
 = \frac{1}{2} \left( \vec{\Phi}_{j} - \vec{\Phi}_{j+1} \right) + \frac{1}{2} \mathcal{K} \left( \vec{\Theta}_{j} + \vec{\Theta}_{j+1} \right)   \; .
\end{split}
\label{eqn:theta-phi-link-multi-component} 
\end{equation}
Note that the on-strip fields $\vartheta^{(I)}_{j^{\ast}}$ are not local in that they contain $\Theta^{(J)}$ on {\em different} layers 
as well as $\Phi^{(I)}$ on the same layer.  

Again in  analogy with the single component case, it is clear that to reproduce the Wen-Zee action \eqref{eqn:def-multi-comp-CS}
we need the following $N_{\text{layer}}$ interactions between the adjacent wires,
\begin{equation}
\mathcal{V} =  \sum_{I=1}^{N_{\text{layer}}} \sum_{j=1}^{N_y} g_{I} \cos \left( 2 \vartheta^{(I)}_{j^{\ast}} \right) \, .
\label{eqn:inter-wire-cos-multicomponent}
\end{equation}
Following the procedure given in Ref.~\cite{Teo-K-14}, we find that, in the strong-coupling limit of $\mathcal{V}$, 
there are $N_{\text{layer}}$ gapless modes at the lower edge $j^{\ast}=0$ described by the following edge Hamiltonian
\begin{equation}
\begin{split}
& \mathcal{H}_{\text{edge}} (j^{\ast}=0) \\
& =
\frac{v^{0}_\text{F}}{2\pi}   \int dx  \sum_{I,J=1}^{ N_{\text{layer}}} 
\left\{ \mathbf{1} + (\mathcal{K}^{-1})^{2} \right\}_{IJ} 
\left( \partial_x \widetilde{\phi}_{0}^{\text{R},(I)}\right)\left( \partial_x \widetilde{\phi}_{0}^{\text{R},(J)}\right)
\end{split}
\label{eqn:edge-Hamiltonian-K}
\end{equation}  
and similarly for the upper edge $j^{\ast}=N_{y}$.     
If we identify the non-universal velocity matrix $\mathbf{V}$ as:
\begin{equation}
\mathbf{V} = 2 v^{0}_\text{F} \left\{ \mathbf{1} + (\mathcal{K}^{-1})^{2} \right\}  \; ,
\end{equation}
the edge Hamiltonian \eqref{eqn:edge-Hamiltonian-K} coincides with that predicted 
by the hydrodynamic Chern-Simons action \eqref{eqn:def-multi-comp-CS} \cite{Wen-95}.   
As has been remarked in Sec.~\ref{sec:topological-action}, that the $K$ matrix, which is of purely topological origin,  
appears in the edge Hamiltonian \eqref{eqn:edge-Hamiltonian-K} through 
the non-universal velocity $\mathbf{V}$ is just an artifact of using the special inter-wire interaction \eqref{eqn:inter-wire-cos-multicomponent} 
to construct the QH state.  

Then, by properly identifying the statistical gauge fields $\{ a^{(I)} \}$, we can derive the Wen-Zee action \eqref{eqn:def-multi-comp-CS} 
with $A=0$.   
Since we follow almost the same steps as before, we just sketch the derivation in Appendix~\ref{sec:WZ-multi-layer}.   
Also it is not hard to use the method in Sec.~\ref{sec:Kohn-mode} to calculate the higher-derivative corrections 
to the topological action \eqref{eqn:def-multi-comp-CS} but we do not give the details here.   
The construction here uses the multi-layer setting.  However, as is described in Appendix \ref{sec:single-layer-const},  
we can always recast the multi-layer system into a single-layer one with {\em longer-range} interactions as in Ref.~\cite{Teo-K-14}.  
\begin{figure}[htb]
\includegraphics[width=60mm]{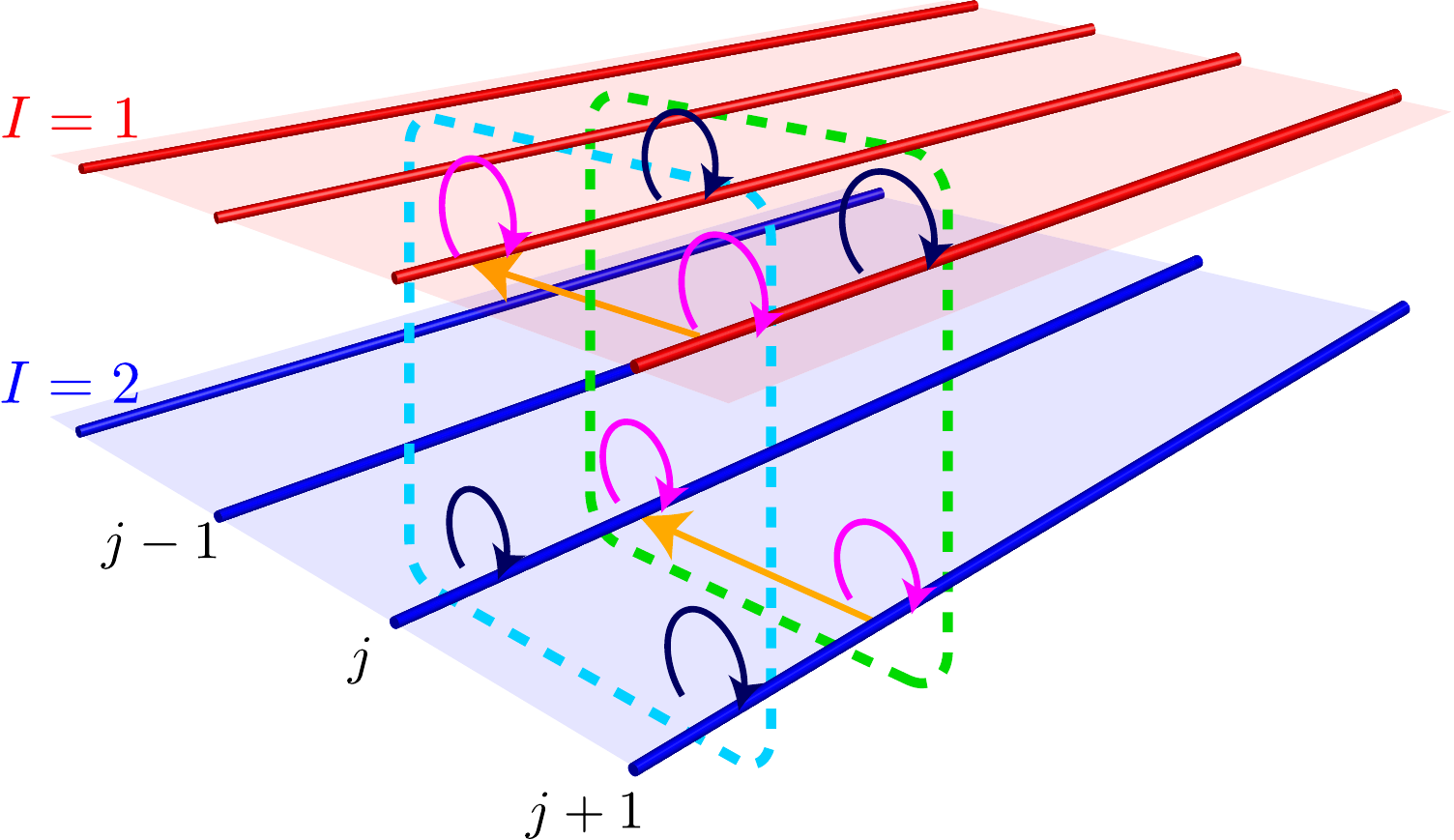} 
\caption{(Color online) %
Multi-layer construction that reproduces the Abelian Chern-Simons gauge theory \eqref{eqn:def-multi-comp-CS}. 
\label{fig:CWC-double-layer2}}
\end{figure}

It is straightforward to rewrite \eqref{eqn:inter-wire-cos-multicomponent} in terms of suitably ordered interactions among fermions or bosons 
that generalize \eqref{eqn:inter-wire-int-by-fermion}.  
In fact, as is shown in Appendix \ref{sec:constraints-on-K}, 
in order for the above interactions to be written only in terms of local operators of the original fermionic/bosonic theories, 
the $N_{\text{layer}}{\times}N_{\text{layer}}$ matrix $K$ must satisfy 
\begin{equation} 
\begin{split}
& \mathcal{K}_{II} = \text{odd} \; , \;\;   \mathcal{K}_{IJ} = \text{even} \; (I \neq J) \quad (\text{for fermions})  \\
&  \mathcal{K}_{IJ} = \text{even}  \;\;  \forall I, J   \quad (\text{for bosons}) \; .
\end{split}
\label{eqn:condition-K}
\end{equation}
Recovering all the Klein factors (in the fermion case) and combining \eqref{eqn:condition-K} with the relation 
$\be^{i \vartheta^{(I)}_{j^{\ast}}} \be^{i \vartheta^{(J)}_{j^{\ast}-1}} 
= (-1)^{\mathcal{K}_{IJ}}  \be^{i \vartheta^{(J)}_{j^{\ast}-1}} \be^{i \vartheta^{(I)}_{j^{\ast}}}$ 
derived from the commutation relations, we see that, for both fermions and bosons, 
all the terms in \eqref{eqn:inter-wire-cos-multicomponent} are commuting and can be minimized simultaneously.  

By requiring translational invariance of the inter-wire interactions, we can determine the filling at which these interactions are 
allowed.  
The magnetic field $b$ and the particle density of each layer may be recovered by the following substitution for all layers $I$ and wires $j$,
\begin{equation}
\begin{split}
\Phi^{(I)}_{j}(x) \to \Phi^{(I)}_{j}(x) + b j x \; , \;\; 
\Theta_{j}^{(I)}(x)  \to \Theta^{(I)}_{j}(x) - \pi \bar{\rho}^{\text{1D}}_{(I)} x  \; .
\end{split}
\label{eqn:recover-b-rho}
\end{equation}
Then, the inter-wire interactions \eqref{eqn:inter-wire-cos-multicomponent} 
acquire additional $x$ dependence $\left( 2\pi \sum_{J} \mathcal{K}_{IJ} \bar{\rho}^{\text{1D}}_{(I)} + b \right) x$ in the cosines.  
The translational invariance requires that these oscillating factors should vanish:
\begin{equation}
\bar{\rho}^{\text{1D}}_{(I)}= \frac{|b|}{2\pi } \sum_{J} (\mathcal{K}^{-1})_{IJ} \; .
\end{equation}
From this, we can read off the corresponding filling fraction as:
\begin{equation}
\nu= \frac{2\pi }{|b|} \sum_{I} \rho^{(I)}_{0} \bar{\rho}^{\text{1D}}_{(I)} =  \sum_{I,J} (\mathcal{K}^{-1})_{IJ}  \; .
\label{eqn:filling-by-K}
\end{equation}
Comparing this with the general expression from the Chern-Simons theory $\nu= \mathbf{q}^{\text{T}} \mathcal{K}^{-1}  \mathbf{q}$, 
we see that our construction corresponds to the symmetric basis $\mathbf{q}=(1,1,\dots,1)$ \cite{Wen-95}.  
Although it is rather straightforward, we will not derive the ground-state degeneracy along the lines in Sec.~\ref{sec:GS-degeneracy} \footnote{%
Now the equivalence relation corresponding to \eqref{eqn:gauge-redundancy-mod-m} is 
\[ \frac{1}{\pi}\sum_{j^{\ast}=1}^{N_y} \delta \vec{\vartheta}_{j^{\ast}} = \sum_{I=1}^{N_{\text{layer}}} n_{I} \mathbf{e}^{(I)}  
\quad (n_{I} \in \mathbb{Z} ) \]
with a set of $N_{\text{layer}}$ vectors defined by $(\mathbf{e}^{(I)})_{J} = K_{JI} =K_{IJ}$.  That is, if the difference $\{ \delta \vec{\vartheta}_{j^{\ast}} \}$ 
between any given pair of ground states is on the lattice spanned by $\{ \mathbf{e}^{(I)} \}$, they are equivalent. Then, the volume of 
a unit cell of this lattice $|\text{det} \, K|$ gives the ground-state degeneracy.},  
since the result also follows directly from the effective Wen-Zee theory \eqref{eqn:def-multi-comp-CS} to be derived 
in Appendix~\ref{sec:WZ-multi-layer} \cite{Vakkuri-W-93,Wesolowski-H-H-94}.  

Regarding the wave functions, there is an important distinction between states formed by multicomponent states 
where the particles are distinguishable by spin or some ``layer'' index \cite{Girvin-M-review-97}, and where they are indistinguishable, 
as in the hierarchy of spin-polarized states in the lowest Landau level.  
In the first case, which includes the Halperin $(m,m,n)$ states \cite{Halperin-83}, one can derive the wave functions 
using the same technique as in Sec.~\ref{sec:Laughlin-wf}.  
The hierarchy wave functions pose a much more difficult problem. Here, the electrons in the  layers 
(or effective Landau levels in the language of composite fermions) are distinguished by their orbital spin.  
This  is a topological quantity that is most directly revealed by geometrical response, since it couples 
to external curvature \cite{Wen-Z-92b}, but it is also related to the Hall viscosity which is a transport coefficient \cite{read2009}.  
It is a challenge to to incorporate the orbital spin in the CWC. 
   
\section{Summary and outlook}
In this paper we have considered the bulk properties in the CWC, i.e., the (topological) ground-state degeneracy, 
the wave function, the bulk effective theory, and the low-energy excitations.   
In the limit of sufficiently strong inter-wire interactions, the ground states are found by minimizing the interaction energy.  
We gave, for the Laughlin states, the precise condition to identify physically inequivalent ground states and showed 
that the CWC correctly reproduces the number of degenerate ground states on a torus.  

In order to describe the low-energy properties in the bulk, we first identified the bosons on the wires as composite bosons and found the expressions 
for the statistical gauge field in the CSGL theory.  These enabled us to obtain the low-energy effective action 
for the statistical (Chern-Simons) gauge field, from which we constructed the bulk Laughlin wave function within the framework of the CWC.  
By integrating out the statistical gauge field, we then derived the hydrodynamic effective action (the Maxwell-Chern-Simons theory) 
in which the leading term is the topological Wen-Zee action.  
We also discussed the effects of anisotropy, which is inherent in the CWC, on the bulk properties.  
The methods developed for the simplest Laughlin states were readily generalized to give a simple recipe for constructing 
general Abelian quantum Hall states, characterized by a $K$ matrix, and the corresponding multi-component Wen-Zee action.  

Our work points at several directions that might be fruitful to explore. 
In the CWC, it is by its nature quite easy to find the gapless edge modes at the boundaries parallel to the wires. 
Although we derived the bulk Wen-Zee action that does encode the chiral edge states at {\em any} boundaries, 
it is not at all clear how this will work microscopically at the boundaries in the direction perpendicular to the wires (i.e., $y$ direction).  
To investigate this, one would have to impose box boundary conditions on the wires but we postpone this for future studies. 

We already mentioned that one can derive the Laughlin quasihole wave functions in the CWC framework, but  that
it is a harder problem to find the wave function for states with quasielectrons, and even harder to find even
the ground state wave functions for the hierarchical states.  
In particular, it is a challenge to understand how the orbital spin, which couples to the curvature of the manifold on which the 
QH liquid is defined, could emerge from a CWC.  

 Another related and difficult, but very interesting, question is whether one could use methods similar to those developed 
 in this paper to extract  topological field theories for non-Abelian states.  
 The obvious first try would be the $\nu =1$ bosonic Moore-Read state for which Teo and Kane found a reasonably simple CWC \cite{Teo-K-14}. 

\underline{\em Note added.} 
Recently, we noticed another very recent related preprint \cite{Fontana-G-H-19}, in which the effective theory
similar to ours has been derived directly without evoking composite bosons. 
We however disagree on how the gauge constraint is handled in this paper.  
\begin{acknowledgments} 
T.H.H. thanks the Yukawa institute for hosting him during the period when this work was carried out, and we thank M.~Hermanns for 
constructive comments on the manuscript. 
Y.I. and K.T. thank K.~Nagao for helpful comments, Y.~Avishai, S.~Brazovski, and B.~Estienne for discussions on related problems, and  
Y.~Fuji for discussions and sharing his unpublished results.  
 K.T. was supported in part by JSPS KAKENHI Grant No.~15K05211 and 18K03455, and T.H.H. by the Swedish Research Council. 
\end{acknowledgments}

\appendix
\section{Ground state degeneracies}
\label{sec:couting-degeneracy}
In Sec.~\ref{sec:CWC-for-Laughlin}, we have claimed that  the coupled-wire system 
with the fine-tuned interaction \eqref{eqn:inter-wire-int-Laughlin} has exactly $m$ different ground states labeled by the value 
\[ 
\sum_{j^{\ast}=1}^{N_y} \bar{\vartheta}_{j^{\ast}} /\pi \;\; (\text{mod $m$}) 
\]
($\vartheta_{0}=\vartheta_{N_y}$ by periodic boundary condition) in the bulk.   In this Appendix we prove this proposition.

The key is that states that are equal up to the periodicity of the bosonic fields must be identified  \cite{Lin-B-F-98,Lecheminant-T-06-SU4}.  
The situation is different for fermionic and bosonic cases as the two cases have different periodic structures (see Fig.~\ref{fig:Gaussian_lattice}).  
The periodicity in the fermionic case is defined with respect to the chiral bosons (the periodicity 
of $(\Phi_j , \Theta_j)$ differs in the fermionic and bosonic sectors): 
\begin{equation}
\phi_{j}^{\text{L}} \sim \phi_{j}^{\text{L}}  + \pi \; , \;\;  
\phi_{j}^{\text{R}}  \sim \phi_{j}^{\text{R}}   + \pi \; .  
\label{eqn:period-F}
\end{equation}
The ``lattice points'' $(\Phi_{j},\Theta_{j})$ identified with $(0,0)$ up to this periodicity are shown in the left panel of Fig.~\ref{fig:Gaussian_lattice}.   
In the bosonic cases, on the other hand, the periodicity is defined as:
\begin{equation}
\Phi_{j} \sim \Phi_{j} + 2\pi \; , \;\; \Theta_{j} \sim \Theta_{j} + \pi 
\label{eqn:period-B}
\end{equation}
and we have a different lattice (see the right panel of Fig.~\ref{fig:Gaussian_lattice}.).  

Let us begin with the  fermion case.  
We take a pair of ground states $\{ \bar{\vartheta}^{(1)}_{j^{\ast}} \}$ and  $\{ \bar{\vartheta}^{(2)}_{j^{\ast}} \}$ 
and denote the difference of the $\bar{\vartheta}^{(1)}_{j^{\ast}} - \bar{\vartheta}^{(2)}_{j^{\ast}}$,  
$\phi_{j}^{(1),\text{L/R}} - \phi_{j}^{(2),\text{L/R}}$, etc. in the two ground states by 
$\delta \vartheta_{j^{\ast}}$, $\delta \phi_{j}^{\text{L/R}}$, etc.   
As $\delta \vartheta_{j^{\ast}} \equiv 0$ (mod $\pi$), it is convenient to specify the difference of the two ground states  
$\{ \delta \vartheta_{j^{\ast}} \}$ by a set of integers $n_{j^{\ast}} \equiv \delta \vartheta_{j^{\ast}} /\pi$.   
Using the definitions 
\[
\begin{split}
& \vartheta_{j^{\ast}} = \frac{1}{2}(\Phi_{j} - \Phi_{j+1}) + \frac{m}{2} (\Theta_{j} + \Theta_{j+1})   
\;\; (j=1,\dots, N_y-1)  \\
& \vartheta_{N_y^{\ast}} = \frac{1}{2}(\Phi_{N_y} - \Phi_{1}) + \frac{m}{2} (\Theta_{N_y} + \Theta_{1})  
=  \vartheta_{0^{\ast}}  \;\; 
(\text{periodic})  \\
& \Phi_{j}  = \phi_{j}^{\text{L}} + \phi_{j}^{\text{R}} \; , \;\;  \Theta_{j}  = \phi_{j}^{\text{L}} - \phi_{j}^{\text{R}}   \; ,
\end{split}
\]
we  readily see that 
\begin{equation}
\sum_{j^{\ast}=1}^{N_y} \delta \vartheta_{j^{\ast}} /\pi = \sum_{j^{\ast}=1}^{N_y}  n_{j^{\ast}}  
= m \sum_{j=1}^{N_y} \left( \delta \phi_{j}^{\text{L}} - \delta \phi_{j}^{\text{R}} \right)/ \pi \; . 
\label{eqn:sum-vartheta-F}
\end{equation}
If the two ground states are equivalent, 
$\delta \phi_{j}^{\text{L}} /\pi, \delta \phi_{j}^{\text{R}}/\pi \equiv 0$ (mod $1$),  
and hence $\sum_{j^{\ast}=1}^{N_y} n_{j^{\ast}} \equiv 0$ (mod $m$).   

Now let us prove the converse, i.e., that if a given pair of ground states satisfy $\sum_{j^{\ast}=1}^{N_y} n_{j^{\ast}} \equiv 0$ (mod $m$), 
then they are equivalent.  To this end, first we show that, by applying a series of  transformations to one of 
the two ground states, we can reduce an arbitrary set $\{n_{1^{\ast}}, n_{2^{\ast}}, \dots n_{N^{\ast}_y}\}$ to a simpler one: 
$\{0,0, \dots 0, M  \}$ with $M = \sum_{i^{\ast}=1} ^{N_y } n_{i^{\ast}}$.    
If we perform the  transformation $\Phi_2 \rightarrow \Phi_2 + 2\pi n_{1^{\ast}}$ to the ground state `1', 
$\delta \vartheta_{1^{\ast}} \rightarrow 0$ in the transformed state, 
leaving us with the new configuration $\{0, n_{1^{\ast}}+n_{2^{\ast}}, n_{3^{\ast}}, \dots n_{N^{\ast}_y}\}$.  
In the next step, we make the transformation $\Phi_3 \rightarrow \Phi_3 -  2\pi (n_{1^{\ast}} + n_{2^{\ast}})$ to obtain 
$\{0, 0,n_{1^{\ast}}+n_{2^{\ast}}+n_{3^{\ast}}, \dots n_{N^{\ast}_y}\}$, and so on.   
After $N_y - 1$ steps, we have got the configuration $\{0,0, \dots 0, M  \}$, where $M = \sum_{i^{\ast}=1} ^{N_y} n_{i^{\ast}} \equiv mk + r $  
with $k \in \mathbb{Z}$ and $r= 0, 1,\dots m-1$.   
Since $\Phi_j \to \Phi_j + 2\pi \mathbb{Z}$ is realizable in both fermionic and bosonic cases, the above procedure is applicable to 
both cases alike.   

Now suppose that $M\equiv 0$ (mod $m$), i.e., we have $\{0, 0, \dots , mk\}$ after the steps described above.   
Then, the question is whether we can find a  transformation that eliminates $mk$ in the last component or not. 
In fact, the transformation depends on the parity of $m$.  
Consider the following two  transformations allowed for fermionic systems [see Eq.~\eqref{eqn:period-F}]: 
$(\phi_{N_y}^{\text{L}},\phi_{N_y}^{\text{R}}) \to (\phi_{N_y}^{\text{L}}  - \pi,\phi_{N_y}^{\text{R}})$  
[$(\Phi_{N_y},\Theta_{N_y}) \to (\Phi_{N_y}- \pi, \Theta_{N_y} - \pi)$] and 
$(\phi_{N_y}^{\text{L}},\phi_{N_y}^{\text{R}}) \to (\phi_{N_y}^{\text{L}},\phi_{N_y}^{\text{R}} - \pi)$ 
[$(\Phi_{N_y},\Theta_{N_y}) \to (\Phi_{N_y} - \pi, \Theta_{N_y} + \pi)$].   They respectively change 
$\{n_1, n_2, \dots n_{N_y}\}$ by
\begin{equation}
\begin{split}
& \{0, 0, \dots 0, -(m-1)/2, -(m+1)/2\}  \\ 
& \text{and} \quad  \{0, 0, \dots 0, (m+1)/2, (m-1)/2\}   \; .
\end{split}
\end{equation}
When $m$ is odd, the above  transformations correctly shift the set $\{ n_j \}$ by integers.  It is clear that if we repeat the first and 
second transformation $(m+1)k/2$ and $(m-1)k/2$ times, respectively, we can reduce 
$\{0, 0, \dots , mk\} \to \{0,0, \dots,0\}$.   Note that the above procedure is allowed only when $m$ is odd, i.e., only for 
the fermionic Laughlin states.    Thus we have proved that a given pair of ground states are equivalent if and only if 
$\sum_{j^{\ast}=1}^{N_y} \delta \vartheta_{j^{\ast}}/\pi \equiv 0$ (mod $m$).   

In the bosonic case, the periodicity is defined by \eqref{eqn:period-B}.   
Since we now have 
\begin{equation}
\sum_{j^{\ast}=1}^{N_y} \delta \vartheta_{j^{\ast}} /\pi = \sum_{j^{\ast}=1}^{N_y}  n_{j^{\ast}} 
= m \sum_{j=1}^{N_y} \delta \Theta_{j} / \pi 
\end{equation}
instead of \eqref{eqn:sum-vartheta-F}, any pair of equivalent ground states ($\delta \Theta_{j} \equiv 0$ mod $\pi$) 
must satisfy the same relation: $\sum_{j^{\ast}=1}^{N_y} n_{j^{\ast}} \equiv 0$ (mod $m$).   
When this relation holds, we can again deform the initial 
$\{n_{1^{\ast}}, n_{2^{\ast}}, \dots n_{N^{\ast}_y}\}$ into $\{0, 0, \dots , mk\}$ by the series of 
 transformations.  Now we consider the two  transformations 
$(\Phi_{N_y},\Theta_{N_y})\to (\Phi_{N_y} -2 \pi,\Theta_{N_y})$ and $(\Phi_{N_y},\Theta_{N_y})\to (\Phi_{N_y} ,\Theta_{N_y} - \pi)$, 
which shift $\{n_{1^{\ast}}, n_{2^{\ast}}, \dots n_{N^{\ast}_y}\}$ by $\{0, 0, \dots 0,1,-1\}$ and  $\{0, 0, \dots 0,- m/2, - m/2\}$, respectively.  
Then, we can eliminate $mk$ by repeating the first and second transformation $mk /2$ and $k$ times, respectively.   
Clearly, this construction requires $m=\text{even}$ ($m/2$ must be integer), which is expected also from the property of the bosonic Laughlin states.  
Therefore, when $m$ is odd (even), the coupled fermionic (bosonic) wires with the inter-wire interaction \eqref{eqn:inter-wire-int-Laughlin} 
exhibit precisely $m$ degenerate ground states. 
A similar but different approach to the ground-state counting based on the edge states has been presented in Ref.~\cite{Sagi-O-S-H-15}.  
\begin{figure}[htb]
\includegraphics[width=0.4\linewidth]{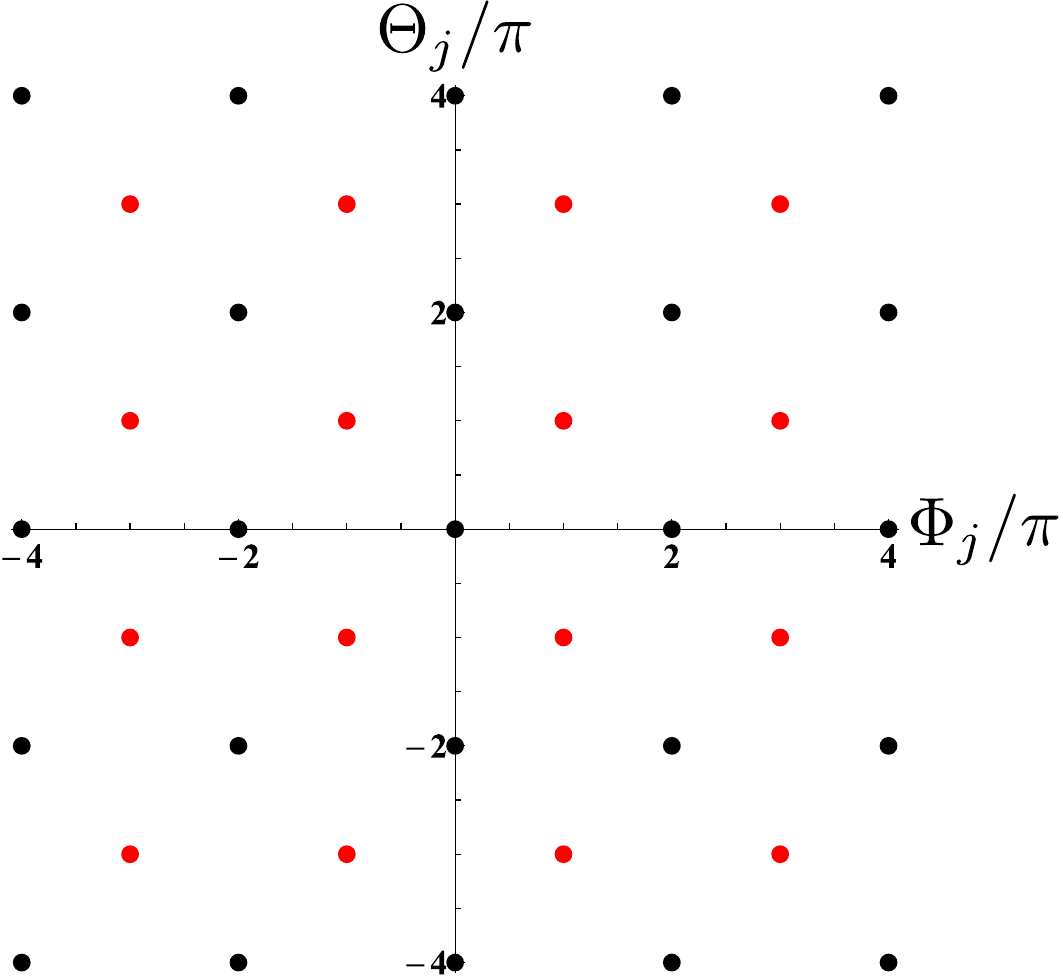} 
\hspace{5mm}
\includegraphics[width=0.4\linewidth]{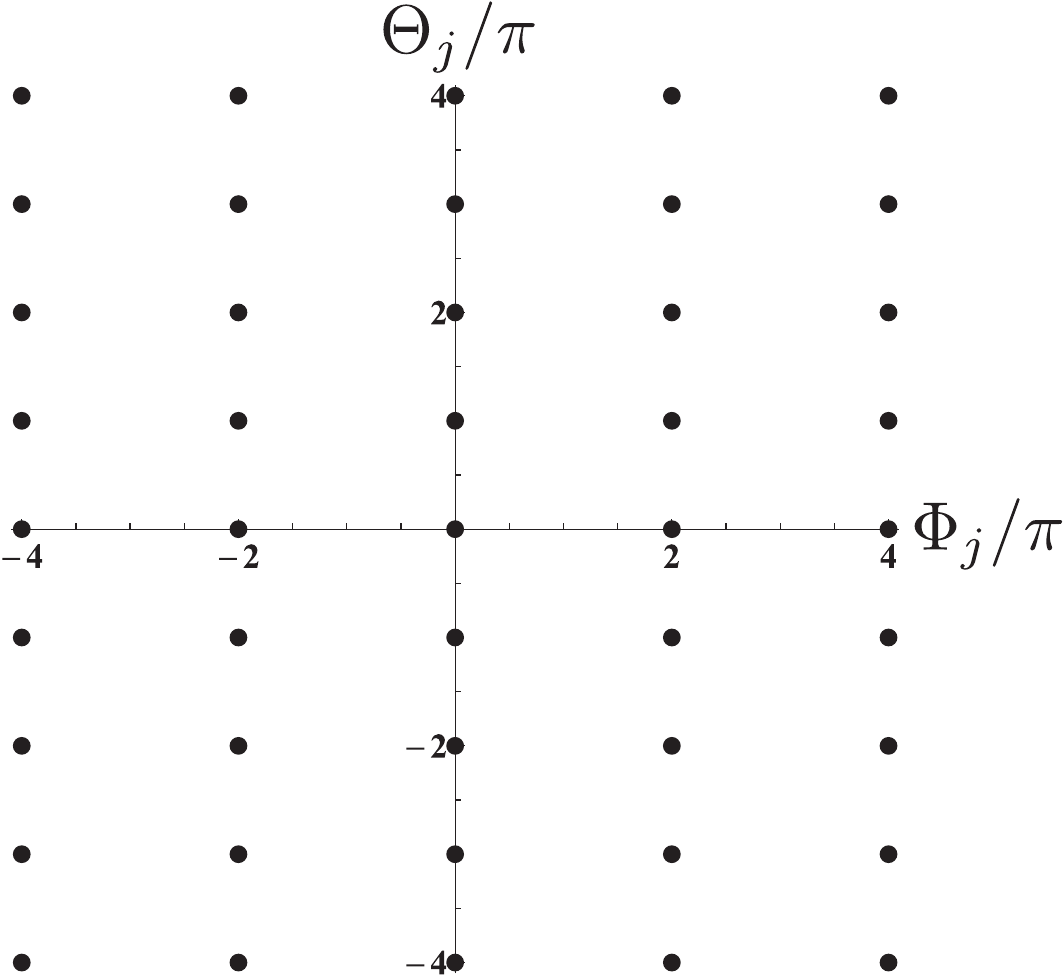} 
\caption{(Color online) %
 Points $(\Phi_j/\pi,\Theta_j/\pi)$ identified by  redundancy for fermionic (left) and bosonic (right) cases.  
 In the fermionic case, the entire lattice splits into two sublattices corresponding to fermionic (red) and bosonic (black) sectors.  
\label{fig:Gaussian_lattice}}
\end{figure}

\section{CWC for general Abelian QH states}
\label{sec:inter-wire-multi-comp-main}
In this Appendix we provide details of the CWC for a general $K$ matrix discussed in Sec.~\ref{sec:general-TQO-2D}, and the 
derivation of the associated Wen-Zee topological field theory.

\subsection{Constraints on $\mathcal{K}$}
\label{sec:constraints-on-K}
So far, we have not assumed any particular statistics of particles on the individual wires.  
However, in order for the above interactions to be written as products of {\em local} operators (say, $R_{j}$ and $L_{j}$) on the wires, 
the elements of the $K$ matrix are constrained.   Any local operators of the wire $(J,j)$ ($J=1,\dots,N_{\text{layer}}$, $j=1,\dots,N_y$) 
can be written as vertex operators of the form:
\begin{equation}
V^{(J)}_{j}(M_j^{(J)},  M_j^{(J)}) = \exp \left\{ 
i M_j^{(J)}\Phi^{(J)}_{j} + i 2 N_j^{(J)}\Theta^{(J)}_{j}   \right\} \; ,
\end{equation}
where the possible values of $(M_j^{(J)},N_j^{(J)})$ are restricted as \cite{Haldane-PRL-81,DiFrancesco-M-S-book}:
\begin{equation}
\begin{split}
& M_j^{(J)} \in \mathbb{Z} \, , \; 2N_j^{(J)}  \in \mathbb{Z} \, , \; (-1)^{M_j^{(J)}+2N_j^{(J)}}=1  \quad \text{for bosons} \\
& M_j^{(J)} \in \mathbb{Z} \, , \;  N_j^{(J)}  \in \mathbb{Z} \quad \text{for bosons} \; .
\end{split}
\label{eqn:selection-rule-M-N}
\end{equation}
The most general expression of the (nearest-neighbor) inter-wire coupling is
\begin{equation}
\begin{split}
& \prod_{J=1}^{N_{\text{layer}}} V^{(J)}_{j}\left( M_j^{(J)}(I),  M_j^{(J)}(I) \right) V^{(J)}_{j+1}\left( M_{j+1}^{(J)}(I),  M_{j+1}^{(J)}(I) \right)   \\
& = \exp \biggl\{ i \sum_{J=1}^{N_{\text{layer}}} \left( M_{j}^{(J)}(I)\Phi^{(J)}_{j} + M_{j+1}^{(J)}(I)\Phi^{(J)}_{j} \right)    \\
& \qquad \qquad 
+ i \sum_{J=1}^{N_{\text{layer}}} \left( 2 N_{j}^{(J)}(I)\Theta^{(J)}_{j} + 2 N_{j+1}^{(J)}(I)\Theta^{(J)}_{j} \right)  \biggr\} \; .
\end{split}
\end{equation}
Comparing this with the $I$-th inter-wire interaction $\cos \left( 2 \vartheta^{(I)}_{j^{\ast}} \right)$, we obtain:
\begin{equation}
\begin{split}
& M_{j}^{(J)}(I) = - M_{j+1}^{(J)}(I) = \delta_{IJ} \\
& 2N_{j}^{(J)}(I) = 2N_{j}^{(J)}(I) = \mathcal{K}_{IJ} 
\end{split}
\end{equation} 
with the integers $\{ M_{j}^{(J)}(I),N_{j}^{(J)}(I)\}$ satisfying the condition \eqref{eqn:selection-rule-M-N}. 
Clearly, for bosonic wires, all the elements $\mathcal{K}_{IJ}$ must be even.  
On the other hand, for systems consisting of fermion wires, 
\begin{equation} 
\mathcal{K}_{II} = \text{odd} \; , \;\;   \mathcal{K}_{IJ} = \text{even} \; (I \neq J) \; .
\end{equation}

The above approach  was was based on having many layers and nearest neighbor interaction between the wires.  
Alternatively, as in Ref.~\cite{Teo-K-14}, we can use a single layer at the expense of having interactions with longer range.  
An example of this is given in Appendix~\ref{sec:single-layer} below.

One can ask if the CWC scheme allows for more general possibilities. A natural extension of the above construction 
is to allow for more than one inter-wire scattering. This can be thought of as forming FQH states of composites of electrons, 
and we provide some details in Appendix~\ref{sec:composite}.    

\subsection{Derivation of multi layer topological action}
\label{sec:WZ-multi-layer}
We now defined a multicomponent statistical gauge field by,
\begin{equation}
\begin{split}
& a^{(I)}_x = -\partial_x \Phi^{(I)}_j \\
&  a^{(I)}_y =  -\frac 2 {l_y} \vartheta^{(I)}_{j^{\ast}}  
\end{split}
\end{equation}
and using the commutation relations for the fields $\vec\Theta_j$ and $\vec\Phi_j$ we get,

\begin{eqnarray}
 [a_{x}^{(I)}(x,j) , a_{y}^{(J)}(x^{\prime},k) ] &=& - 2\pi i K_{IJ} \left( \frac{1}{l_{y}} \delta_{jk} \right) \delta(x-x^{\prime}) \nonumber  \\
&\to & - 2\pi i K_{IJ}  \delta(\vec{x}-\vec{x}^{\prime}) \,
\end{eqnarray}
These commutation relations amounts to having an action with the kinetic term,
\begin{equation}
S = \frac{1}{2\pi}\int\! d^{3} x  \sum_{I,J} \mathcal{K}_{IJ}^{-1} \epsilon^{ij} {a}_{i}^{(I)}(x) \dot  a_{j}^{(J)}(x)  \; .  \label{multikin}
\end{equation}
As in Sec.~\ref{sec:CSGL}, the corresponding Hamiltonian is given by:
\begin{equation}
\mathcal H = \sum_I \left [\beta_x^I (a_x^{(I)})^2 + \beta_y^I (a_y^{(I)})^2 \right] \, , 
\label{eqn:multiham}
\end{equation}
where the coefficients $\beta_i^I$ are easily extracted from the  Luttinger Hamiltonians for the individual wires, 
and the expansion of the inter-wire interactions \eqref{eqn:inter-wire-cos-multicomponent}.  
With \eqref{multikin} and \eqref{eqn:multiham} in hand, we can step by step follow the derivation in Sec.~\ref{sec:hydrodynamic-WZ} 
to arrive at the Wen-Zee Lagrangian \eqref{eqn:def-multi-comp-CS} with a charge vector ${\mathbf q} = (1,1,\dots 1)$ 
corresponding to the symmetric basis of the $K$ matrix using the terminology of Wen \cite{Wen-95}.  
Using the above charge vector $\mathbf{q}$, we can write the filling fraction $\nu$ compactly as 
$\nu= \mathbf{q}^{\text{T}} \mathcal{K}^{-1}  \mathbf{q}$ reproducing the well-known formula. 

Integrating the statistical gauge fields $a_i^I$ in the presence of the Hamiltonian \eqref{eqn:multiham} 
will generate higher derivative corrections to to the topological action \eqref{eqn:def-multi-comp-CS} 
and the resulting hydrodynamical theory can be used to 
study collective modes, just as in the single component case.

\subsection{Single-layer description} \label{sec:single-layer}
\label{sec:single-layer-const}
Above we gave  the CWC  for generic Abelian topological states using the multi-layer scheme, 
where we used $N_{\text{layer}}=n$ sheets of coupled wires to realize topological states characterized by an $n$-dimensional $K$ matrix.   
However, we can easily transform the multi-layer scheme to a single-layer one proposed 
in, e.g., Ref.~\onlinecite{Teo-K-14} by ``crushing'' the stack of layers.  The idea is to first relabel 
the $j$-th wire on the $I$-th layer $(I,j)$ ($I=1,\ldots, n$, $j=1,\ldots, N_{y}$) as the $j^{\prime}$-th one [$j^{\prime}=n(j-1)+I$] 
on a {\em single} layer.   
Now the interaction between the wires $(I,j)$ and $(J,j+1)$ is transformed to a long-range one [with the range-$(n+J-I)$] between 
the wires $n(j-1)+I$ and $nj+J$.  
This way, we formally rewrite the original $n$-layer system in terms of a {\em single}-layer system 
including at most range-$(2n-1)$ interactions.  However, this is not the end of the story.  In fact, there is a freedom 
of changing the boson fields while preserving the commutation relations.  

Let us demonstrate how the procedure described above works 
in the Haldane-Halperin $\nu=2/5$ state \cite{Haldane-FQH-83,Halperin-hierarchy-84} characterized by the 
following $K$ matrix (in the symmetric basis)
\begin{equation}
\mathcal{K} = 
\begin{pmatrix}
3 & 2 \\ 2 & 3 
\end{pmatrix} \; .
\end{equation}
According to the procedure described in Sec.~\ref{sec:general-TQO-2D} [see Eq.~\eqref{eqn:modified-chiral-multi-comp}], 
the modified chiral bosons are defined as:
\begin{subequations}
\begin{align}
\begin{split}
& \tilde{\phi}_{j}^{\text{L}(1)}= (\Phi_{2j-1} + 3 \Theta_{2j-1} +2\Theta_{2j} )/2   \\
& \tilde{\phi}_{j}^{\text{R}(1)}= (\Phi_{2j-1} -3 \Theta_{2j-1} -2\Theta_{2j} )/2 
\end{split}
\label{eqn:modified-boson-2layer-a}
\\
\begin{split}
& \tilde{\phi}_{j}^{\text{L}(2)}= (\Phi_{2j} +2 \Theta_{2j-1} + 3 \Theta_{2j} )/2   \\
& \tilde{\phi}_{j}^{\text{R}(2)}= (\Phi_{2j} -2 \Theta_{2j-1} -3 \Theta_{2j} )/2  \; ,
\end{split}
\label{eqn:modified-boson-2layer-b}
\end{align}
\end{subequations}
where we have made the replacement: 
$(\Phi^{(1)}_{j},\Theta^{(1)}_{j},\Phi^{(2)}_{j},\Theta^{(2)}_{j}) \to (\Phi_{2j-1},\Theta_{2j-1},\Phi_{2j},\Theta_{2j})$ 
($j=1,\ldots,N_{y}$) on the right-hand side.  
In terms of the new set of variables, the original inter-wire interactions read as
\begin{equation}
\begin{split}
& \cos \left[ 2 \vartheta^{(1)}_{j^{\ast}} \right] = \cos \left[ 2(\widetilde{\phi}{}^{\text{L}(1)}_{j}  -  \widetilde{\phi}{}^{\text{R}(1)}_{j+1} ) \right]  \\
& = \cos \bigl\{ 
\left(\Phi_{2j-1}-\Phi_{2j+1} \right) +3 \Theta_{2j-1}+2\Theta_{2j}  \\
& \qquad  
+ 3 \Theta_{2j+1} + 2\Theta_{2j+2}  \bigr\}  \\
& \cos \left[ 2 \vartheta^{(2)}_{j^{\ast}} \right] = \cos \left[ 2(\widetilde{\phi}{}^{\text{L}(2)}_{j}  -  \widetilde{\phi}{}^{\text{R}(2)}_{j+1} ) \right]  \\
& = \cos \bigl\{ 
\left(\Phi_{2j-1}-\Phi_{2j+1} \right) +2 \Theta_{2j-1}+3\Theta_{2j}  \\
& \qquad + 2 \Theta_{2j+1} + 3\Theta_{2j+2} ) \bigr\}  \; ,
\end{split}
\end{equation}
which include backscattering processes on four wires $(2j-1,2j,2j+1,2j+2)$ as well as single-particle hoppings between 
second-neighbor wires.  Therefore, if we squeeze the double-layer systems to a single-layer one, interactions involving four wires 
are introduced.

Now we show that we can reduce the number of wires involved in the inter-wire interactions by the redefinition of 
the chiral bosons in Eqs.~\eqref{eqn:modified-boson-2layer-a} and \eqref{eqn:modified-boson-2layer-b}.  
In fact, we can readily check that all the commutation relations among $\{ \tilde{\phi}_{j}^{\text{L/R}(I)} \}$ 
are preserved even after we redefine the chiral bosons as: 
\begin{equation}
\tilde{\phi}_{j}^{\text{L/R}(1)} \to \tilde{\phi}_{j}^{\text{L/R}(1)} + \Theta_{j}^{(2)} \; , \;\;
\tilde{\phi}_{j}^{\text{L/R}(2)} \to \tilde{\phi}_{j}^{\text{L/R}(2)} - \Theta_{j}^{(1)} \; .
\end{equation}
Then, it is clear that all the arguments on the underlying topological properties in Sec.~\ref{sec:WZ-multi-layer} carry over 
and that the same topological phase is obtained for the new system as well.   
With the new strip variables defined by $\vartheta_{j^{\ast}}^{(I)}=\tilde{\phi}_{j}^{\text{L/R}(I)}-\tilde{\phi}_{j+1}^{\text{L/R}(I)}$, 
the two inter-wire interactions now read as
\begin{equation}
\begin{split}
& \cos \left[ 2 \vartheta^{(1)}_{j^{\ast}} \right]  \\
&= \cos \left\{  \left( \Phi^{(1)}_{j} - \Phi^{(1)}_{j+1} \right) + 3\left( \Theta^{(1)}_{j} + \Theta^{(1)}_{j+1} \right) + 4 \Theta^{(2)}_{j} \right\} \\
& \cos \left[ 2 \vartheta^{(2)}_{j^{\ast}} \right]  \\
& = \cos \left\{  \left( \Phi^{(2)}_{j} - \Phi^{(2)}_{j+1} \right) + 4 \Theta^{(1)}_{j+1} + 3\left( \Theta^{(2)}_{j} + \Theta^{(2)}_{j+1} \right)  \right\}  \; .
\end{split}
\end{equation}
After relabeling the wires as before, we obtain the interactions proposed in Ref.~\cite{Teo-K-14} containing only three-wire couplings.  

\subsection{FQHE of composite particles} \label{sec:composite}
So far, we have been discussing the case only with single-particle (inter-wire) hopping where the coefficients of the $\Phi$ fields 
appearing in the inter-wire interactions are always $\pm 1$.  
However, we may think of the situations where multi-particle hopping occurs, or more specifically, when we have 
the following inter-wire interactions:
\begin{equation}
\sum_{I=1}^{N_{\text{layer}}} \sum_{j=1}^{N_y} 
g_{I} \cos \left\{ n \left( \Phi^{(I)}_{j} - \Phi^{(I)}_{j+1} \right)  
+ \sum_{J} \widetilde{\mathcal{K}}_{IJ} \left( \Theta^{(J)}_{j} + \Theta^{(J)}_{j+1} \right) \right\}  \nonumber
\; .
\label{eqn:inter-wire-cos-multicomponent-2}
\end{equation}
In the above, $\widetilde{\mathcal{K}}$ is an invertible, symmetric integer-valued matrix 
which is {\em not} necessarily the $K$ matrix of the underlying Chern-Simons theory.  
As in the previous cases, we can formally introduce the following chiral bosons:
\begin{equation}
\vec{\widetilde{\phi}}{}^{\text{L}}_{j} \equiv \frac{1}{2} \left( n \vec{\Phi}_{j} 
+ \widetilde{\mathcal{K}} \vec{\Theta}_{j} \right)  \; , \;\;
\vec{\widetilde{\phi}}{}^{\text{R}}_{j} \equiv \frac{1}{2} \left( n \vec{\Phi}_{j} 
- \widetilde{\mathcal{K}} \vec{\Theta}_{j} \right)  \; .
\label{eqn:modified-chiral-multicomponent-3}
\end{equation}
Then, the on-strip fields defined by:
\begin{equation}
\begin{split}
& \vec{\varphi}_{j^{\ast}} \equiv \vec{\widetilde{\phi}}{}^{\text{L}}_{j}  +  \vec{\widetilde{\phi}}{}^{\text{R}}_{j+1} 
= \frac{n}{2} \left( \vec{\Phi}_{j} + \vec{\Phi}_{j+1} \right) + \frac{1}{2} \widetilde{\mathcal{K}} \left(  \vec{\Theta}_{j} - \vec{\Theta}_{j+1} \right)  \\ 
& \vec{\vartheta}_{j^{\ast}} \equiv \vec{\widetilde{\phi}}{}^{\text{L}}_{j}  -  \vec{\widetilde{\phi}}{}^{\text{R}}_{j+1}  
= \frac{n}{2} \left( \vec{\Phi}_{j} - \vec{\Phi}_{j+1} \right) + \frac{1}{2} \widetilde{\mathcal{K}} \left(  \vec{\Theta}_{j} + \vec{\Theta}_{j+1} \right)    
\end{split}
\label{eqn:theta-phi-link-multi-component-2}
\end{equation}
enable us to rewrite the above inter-wire interactions as:
\begin{equation}
\sum_{I=1}^{N_{\text{layer}}} \sum_{j=1}^{N_y} g_{I} \cos \left( 2 \vartheta^{(I)}_{j^{\ast}} \right)   \; .
\end{equation}
The hidden symmetry now reads:
\begin{equation}
\begin{pmatrix}
\vec{\widetilde{\phi}}{}^{\text{L}}_{j} \\
\vec{\widetilde{\phi}}{}^{\text{R}}_{j+1}
\end{pmatrix} 
\to 
\begin{pmatrix}
\vec{\widetilde{\phi}}{}^{\text{L}}_{j}   \\
\vec{\widetilde{\phi}}{}^{\text{R}}_{j+1} 
\end{pmatrix}
+  \frac{n}{2} 
\begin{pmatrix}  \widetilde{\mathcal{K}} \vec{\chi}_{j^{\ast}} \\ \widetilde{\mathcal{K}} \vec{\chi}_{j^{\ast}} \end{pmatrix}
\; ,
\label{eqn:U1-residual-multicomponent-2}
\end{equation}
with the $N_{\text{layer}}$-dimensional vector $\vec{\chi}_{j^{\ast}}$ parametrizing the residual U(1)$^{N_{\text{layer}}}$ symmetry.  

\begin{equation}
\begin{split}
& a^{(I)}_{x}(x,j) \equiv \sum_{J} (\widetilde{\mathcal{K}}^{-1})_{IJ} \partial_{x} \Phi^{(J)}_{j}(x)   \\
& a^{(I)}_{y}(x,j) \equiv \frac{2}{n} \frac{1}{l_{y}}  \Theta^{(I)}_{j}(x)  \; .
\end{split}
\label{eqn:a-by-bosons-multicomponent-2b}
\end{equation}
Plugging these expressions into the Berry-phase part of the Luttinger-liquid action, we obtain:
\begin{equation}
- \int\! d^{3}x \frac{1}{4\pi} \sum_{I,J} \left( n \widetilde{\mathcal{K}}_{IJ} \right)  \epsilon^{ij} \dot{a}_{i}^{(I)}(x)  a_{j}^{(J)}(x)  \; ,
\label{eqn:BP-to-CS-2}
\end{equation}
which immediately implies that the underlying topological field theory is the Wen-Zee action 
with the $K$ matrix given by $n\widetilde{\mathcal{K}}$.   
\begin{equation}
\begin{split}
\tilde{\rho}^{(I)}_{j^{\ast}} (x) & 
= \frac{1}{\pi} \sum_{J=1}^{N_{\text{layer}}} (\widetilde{\mathcal{K}}^{-1})_{IJ} \partial_{x} \vartheta^{(J)}_{j^{\ast}} (x)  \\
& \simeq 
\frac{n}{2\pi} l_{y}  \left( \partial_{x} a_{y}^{(I)} - \partial_{y} a_{x}^{(I)} \right) \; .
\end{split}
\label{eqn:density-by-A-2b}
\end{equation}
%
\bibliographystyle{apsrev4-1}
\bibliography{./references/CWC_bibliography,./references/misc,./references/field_theory,./references/QHE-TI,./references/topological_order,./references/ref}
\end{document}